\documentclass{article}

\usepackage{arxiv}

\usepackage[utf8]{inputenc} 
\usepackage[T1]{fontenc}    
\usepackage{hyperref}       
\usepackage{url}            
\usepackage{booktabs}       
\usepackage{multirow}       
\usepackage{amsfonts}       
\usepackage{amsthm}         
\usepackage{nicefrac}       
\usepackage{microtype}      
\usepackage{lipsum}		
\usepackage{graphicx}
\usepackage[numbers]{natbib}
\usepackage{doi}
\usepackage{xcolor}
\usepackage{amsmath}
\usepackage{amsthm}

\newtheorem{theorem}{Theorem}
\newtheorem{assumption}{Assumption}
\newtheorem{corollary}{Corollary}
\usepackage{dsfont}
\usepackage{algorithm}
\usepackage{algpseudocode}
\usepackage{setspace}
\usepackage{graphicx}
\usepackage{subcaption}

\title{Analytical Formula for Fractional-Order Conditional Moments of Nonlinear Drift CEV Process with Regime Switching: Hybrid Approach with Applications}

\author{ \href{https://orcid.org/0000-0002-9186-6891}{\includegraphics[scale=0.06]{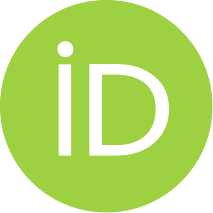}\hspace{1mm}Kittisak Chumpong}\\
	Division of Computational Science\\
	Faculty of Science, Prince of Songkla University\\
	Songkhla 90110, Thailand \\
	\texttt{kittisak.ch@psu.ac.th} \\
	\And
	\href{https://orcid.org/0000-0001-7797-4418}{\includegraphics[scale=0.06]{orcid.pdf}\hspace{1mm}Khamron Mekchay}\thanks{Corresponding author}  \\
	Department of Mathematics and Computer Science \\
	Faculty of Science, Chulalongkorn University\\
	 Bangkok 10330, Thailand \\
	\texttt{khamron.m@chula.ac.th} \\
	\And
	\href{https://orcid.org/0000-0002-6395-6005}{\includegraphics[scale=0.06]{orcid.pdf}\hspace{1mm}Fukiat Nualsri}  \\
	Department of Mathematics and Computer Science \\
	Faculty of Science, Chulalongkorn University\\
	 Bangkok 10330, Thailand \\
	\texttt{6373011023@student.chula.ac.th} \\
	\And
	\href{https://orcid.org/0000-0002-0986-2518}{\includegraphics[scale=0.06]{orcid.pdf}\hspace{1mm}Phiraphat Sutthimat}  \\
	Department of Mathematics \\
	Faculty of Science, Kasetsart University\\
	 Bangkok 10900, Thailand \\
	\texttt{phiraphat.sut@ku.th} \\ 
}

\renewcommand{\shorttitle}{Analytical Formula for Fractional-Order Moments in NLD-CEV with Regime Switching}

\hypersetup{
pdftitle={Fractional Conditional Moments of NLD-CEV with Regime Switching},
pdfsubject={Conditional moments},
pdfauthor={Kittisak Chumpong, Khamron Mekchay, Fukiat Nualsri, Phiraphat Sutthimat},
pdfkeywords={Nonlinear drift CEV process, Regime switching, Markov chain, Markovian switching, Hybrid system, VIX futures},
}

\begin{document}
\maketitle

\begin{abstract}
    This paper presents the development of an analytical methodology for computing fractional-order conditional moments in nonlinear drift constant elasticity of variance (NLD-CEV) processes, where regime transitions are governed by continuous-time finite-state irreducible Markov chains. Through the implementation of a hybrid systems framework, we establish closed-form solutions for conditional moments spanning arbitrary fractional orders across multiple regime states, significantly advancing the analytical tractability of NLD-CEV processes under stochastic regime conditions. The theoretical foundation of our approach relies on the construction and resolution of an intricate system of coupled partial differential equations, derived through the application of the Feynman--Kac formula in the context of switching diffusions. 
    Our analysis extends to a detailed examination of asymptotic behaviors exhibited by fractional-order conditional moments within a two-state regime-switching framework, particularly emphasizing the interplay between the Markov chain intensity matrix\rq s symmetry properties and various parametric configurations in determining the process' evolution. To illustrate the practical relevance of our approach, Monte Carlo simulations for the process with regime-switching are applied to validate the accuracy and computational efficiency of the analytical formulas. Furthermore, to demonstrate significant improvements over traditional methods, we apply our findings for the valuation of financial derivatives within a dynamic nonlinear mean-reverting regime-switching process. This work offers substantial contributions to financial modeling and derivative pricing by providing a robust tool for practitioners and researchers who are dealing with complex stochastic~environments.
\end{abstract}


\section{Introduction}

In financial modeling, accurately capturing the stochastic behavior of asset prices is crucial for assessing risk, pricing derivatives, and guiding investment strategies. Standard models, such as the Black--Scholes model, while revolutionary, assume constant volatility and linear dynamics, limiting their ability to address real-world complexities, especially under volatile or turbulent market conditions~\mbox{\cite{carr2017leverage,dumas1998implied}}. To better address these dynamics, researchers have turned to stochastic processes with flexible variance structures and regime-switching mechanisms, which allow models to adapt for shifting economic conditions and unexpected market events~\cite{jones2003nonlinear}.

Among these advances, models incorporating a constant elasticity of variance (CEV) have gained prominence for their ability to capture changing volatility patterns that depend on the asset price itself~\mbox{\cite{cox1975notes,cox1976valuation,jones2003nonlinear}}. The CEV model introduces a degree of elasticity that makes the volatility proportional to the level of the underlying asset, capturing the empirically observed the leverage effect seen in financial markets. However, while the CEV model has made critical improvements, it remains limited in handling nonlinear dynamics and varying market conditions over time~\cite{ahn1999parametric}.

The nonlinear drift constant elasticity of variance (NLD-CEV) process builds on the flexibility of the CEV model, integrating a nonlinear drift component and regime-switching capability to adapt dynamically across market environments~\cite{marsh1983stochastic}. This process is particularly suited for capturing the asymmetries and conditional heteroscedasticity that are characteristic of asset returns, where market regimes fluctuate according to changing economic conditions \mbox{\cite{ahn1999parametric,anderson2012continuous,chumpong2024closed}}. The NLD-CEV process incorporates continuous-time finite-state Markov chains, allowing the process' parameters to transition between different states. This regime-switching mechanism enables the model to adjust dynamically, accounting for sudden changes in volatility or drift, which are common in markets influenced by external shocks or cyclical shifts \mbox{\cite{boyarchenko2009american,zhu2012new}}.

The application of regime-switching models has become instrumental in derivative pricing across diverse market frameworks. A significant advancement came from Buffington and Elliott~\cite{buffington2002regime}, who enhanced the Black--Scholes framework by incorporating regime-dependent variables, including interest rates, drift, and volatility. Their work yielded the characteristic function for occupation times and adapted the Barone--Adesi--Whaley approximation, thus facilitating a more accurate valuation of both European and American options. A notable computational breakthrough emerged through Zhu, Badran, and Lu~\cite{zhu2012new}, who developed an elegant analytical solution for European option pricing within a~two-state regime-switching economy. Their innovation transformed a computationally intensive double integral into a more manageable single integral of elementary functions, substantially enhancing both computational efficiency and precision, in line with broader purely numerical advances in the solution of option pricing partial differential equations (PDEs)~\cite{liu2024solving}. The application domain expanded further when Li~\cite{li2016trading} developed a regime-switching framework addressing optimal volatility index (VIX) futures trading. This approach utilized the Cox--Ingersoll--Ross (CIR) process coupled with a finite-state Markov chain to capture the market dynamics, enabling traders to optimize their entry and exit strategies through solutions derived from variational inequalities. These developments were subsequently enriched by Lin and He~\cite{lin2020regime}, who investigated European option pricing through a fractional Black--Scholes model incorporating regime switching. Their methodology captured the essential characteristics of asset returns through a sophisticated two-step solution procedure for the coupled fractional PDE system, culminating in an exact pricing formula based on Fourier cosine series expansion that demonstrated both rapid convergence and practical applicability. For comprehensive coverage of additional models and recent developments in regime-switching applications, the readers may consult \mbox{\cite{elliott2013pricing,he2023new,shen2013pricing}}.

The proposed framework is motivated by the limitations of previous models, such as the extended~CIR (ECIR) process~\cite{hull1990pricing}, which, while incorporating mean-reversion and regime-switching dynamics, relies primarily on linear drift components and lacks the elasticity offered by the NLD-CEV approach. In the ECIR process, the model captures conditional moments through a system of PDEs under Markov switching, enabling applications in VIX futures pricing. However, the ECIR\rq s linearity may limit its application in scenarios requiring greater adaptability to changing market volatility. The NLD-CEV process addresses this gap by introducing a nonlinear drift term, which allows for more robust modeling of asymmetrical market responses and variance elasticity. Compared with existing regime-switching NLD-CEV-type models, which typically allow regime dependence only in a limited subset of model parameters, the proposed framework further generalizes the model by allowing full regime dependence across its key structural components. This full regime dependence leads to a coupled hybrid PDE system for fractional-order conditional moments, for which explicit closed-form solutions are derived in this paper.

In this paper, we develop an analytical framework to calculate fractional-order conditional moments for the NLD-CEV process with regime switching. Using a hybrid system approach, we derive closed-form solutions for these moments, enhancing both the analytical tractability and computational efficiency of the model. Our approach is grounded in the Feynman--Kac formula, adapted for regime-switching diffusion processes, leading to a recursive system of PDEs for the fractional moments. We also explore the model\rq s asymptotic properties in a two-state regime-switching scenario, assessing how variations in the Markov chain intensity matrix and parameter configurations affect the conditional moments and overall behavior of the process. This analysis demonstrates the ability of the NLD-CEV process to handle a wide range of market scenarios, extending its applicability beyond that of traditional~models.

To validate the efficacy of our approach, we conduct Monte Carlo (MC) simulations, comparing the performance of our closed-form solutions with traditional computational methods. The results illustrate not only the accuracy but also the computational efficiency of our analytical formulas in capturing the conditional moments. Moreover, we apply the NLD-CEV process for the valuation of financial derivatives, such as options, highlighting its practical relevance and superiority over traditional models in capturing market complexities. Our findings present a valuable contribution to the fields of financial modeling and derivative pricing, equipping researchers and practitioners with a powerful tool for navigating stochastic environments characterized by nonlinear dynamics and regime-switching~behavior.

The paper is organized as follows. In Section~\ref{sec_Regime}, we define the NLD-CEV process with regime switching, setting up the necessary assumptions that support our model. Section~\ref{sec_CM} delves into the hybrid system of PDEs derived from the Feynman--Kac formula for switching diffusions, laying out the analytical foundation for computing fractional-order conditional moments under a regime-switching structure. Moreover, we present the main analytical results, including explicit closed-form formulas for conditional moments, and examine the influence of different parameter configurations and Markov chain states. Section~\ref{sec_exper} provides a method for stochastic differential equations (SDEs) with regime switching and validates our theoretical findings through MC simulations, comparing the efficiency and accuracy of formulas against conventional numerical methods. Finally, Section~\ref{sec_application} discusses the practical applications of the NLD-CEV process in financial contexts, particularly in pricing derivatives such as VIX futures, illustrating the model\rq s adaptability to real-world problems. Section~\ref{sec_conclusion} concludes the paper with a summary of our findings and potential directions for future~research.

\section{The $m$-state regime-switching NLD-CEV process} \label{sec_Regime}

The CEV diffusion process, first introduced by Cox~\cite{cox1975notes} in 1975, is viewed as an extension of the Ornstein--Uhlenbeck (OU) process, particularly for applications in finance. Since its inception, the~CEV model has been explored and expanded across various fields. In recent studies, Araneda et al.~\cite{araneda2021sub} investigated the sub-fractional CEV model, while Cao et al.~\cite{cao2021pricing} examined variance swap pricing under a hybrid CEV--stochastic volatility process, demonstrating the model\rq s adaptability to contemporary financial challenges.


\subsection{The nonlinear drift constant elasticity of variance process}
\medskip
The first generalized version of Cox\rq s CEV process was proposed by Marsh and Rosenfeld~\cite{marsh1983stochastic}, who incorporated time-dependent parameters and a nonlinear drift term, resulting in what we now call the NLD-CEV process, as described in~\cite{sutthimat20222closed}. The NLD-CEV process can be formulated as
\begin{equation} \label{main_process}
    dR_t = \kappa(t) \Big(\theta(t)R_t^{-(1-\beta)} - R_t \Big) dt + \sigma(t)R_t^{\beta/2} dW_t^\mathbb{Q}, \quad \beta \in [0,2) \cup (2, \infty),
\end{equation}
where $\theta(t)$, $\kappa(t)$, and $\sigma(t)$ are time-dependent parameters over $t \in [0, T]$ with the initial value $R_{0} > 0$, and~$W_t^\mathbb{Q}$ is a standard Wiener process under the probability space $(\Omega,\mathcal{F},\mathbb{Q})$ with filtration $(\mathcal{F}_{t})_{t\geq0}$  and a risk-neutral probability measure $\mathbb{Q}$. Here, the diffusion term $\sigma(t)R_t^{\beta/2}$ aligns with the structure in Cox\rq s original CEV model, but the drift term $\kappa(t)\big(\theta(t)R_t^{-(1-\beta)} - R_t\big)$ introduces nonlinearity, distinguishing it from the traditional formulation. We restrict our attention to $\beta \ge 0$ in \eqref{main_process}. For $\beta<0$, the diffusion-induced random fluctuations become excessively strong near the zero boundary, substantially increasing the risk that $R_t$ crosses zero and reaches negative values. Such behavior contradicts the intrinsic non-negativity required for many financial quantities, such as asset prices or interest rates, and is therefore excluded from the present analysis.

Varying the parameter $\beta$ produces several well-known processes within the NLD-CEV process. When $\beta = 1$, the model becomes the ECIR process; for $\beta = 0$, it aligns with the OU process; for~$\beta \rightarrow 2$, it approximates the lognormal process studied by Merton; and when $\beta = 3$, it corresponds to the inverse Feller (IF) or $3/2$~stochastic volatility model (SVM).

Evidence suggests that nonlinearity in the drift component of the NLD-CEV process is particularly well-suited for modeling the behavior of financial derivatives, especially those influenced by interest~rate dynamics~\cite{zhou2003ito}. For instance, the empirical findings of Chan et al. \cite{chan1992empirical} demonstrate that models with~$\beta >2$ more effectively capture short-term rate movements compared with models where $\beta < 2$. The~NLD-CEV process is divided into two cases, each characterized by a distinct range of values for~$\beta$. In the first case, where $0 \leq \beta < 2$, we set $\beta = \frac{2\alpha - 1}{\alpha}$, and the process becomes
\begin{equation} \label{main_process<2}
    dR_t = \kappa(t) \Big( \theta(t)R_t^{\frac{\alpha - 1}{\alpha}} - R_t \Big) dt + \sigma(t)R_t^{\frac{2\alpha - 1}{2\alpha}} dW_t^\mathbb{Q}, \quad \alpha \geq \frac{1}{2}.
\end{equation}
To guarantee the existence of a unique pathwise strong solution for~\eqref{main_process<2}, the two following existence and uniqueness conditions are required.

\begin{assumption} \label{Assumption1}
    The parameters $\kappa(t), \theta(t)$, and $\sigma(t)$ in the NLD-CEV process~\eqref{main_process} are strictly positive and smooth functions depending on the temporal variable $t\in[0,T]$. Moreover, ${\kappa(t)}/{\sigma^{2}(t)}$ is locally bounded on $[0,T]$.
\end{assumption}

\begin{assumption} \label{Assumption2}
    The process $R_t$ in~\eqref{main_process} contains the inequality $2\kappa(t)\theta(t) \geq \sigma(t)^2$.
\end{assumption}

 Under Assumptions~\ref{Assumption1} and \ref{Assumption2}, the drift and diffusion coefficients of \eqref{main_process<2} satisfy standard local Lipschitz continuity and appropriate growth conditions on the admissible state domain. In particular, the imposed regularity and positivity of the model parameters ensure that the coefficients are well-defined and locally bounded, while the structural condition in Assumption~\ref{Assumption2} prevents the degeneracy of the diffusion term. As a result, the SDE \eqref{main_process<2} admits a unique strong (pathwise unique) solution up to the given time horizon; see \cite{chumpong2022simple}.

In the second case, where $\beta > 2$, we set $\beta = \frac{2\alpha + 1}{\alpha}$, resulting in
\begin{equation} \label{main_process>2}
    dR_t = \kappa(t) \Big( \theta(t)R_t^{\frac{\alpha + 1}{\alpha}} - R_t \Big) dt + \sigma(t)R_t^{\frac{2\alpha + 1}{2\alpha}} dW_t^\mathbb{Q}, \quad \alpha > 0.
\end{equation}
It is noteworthy that as $\beta \rightarrow 2$, both cases of the NLD-CEV process converge, with $\alpha \rightarrow \infty$ in each scenario, linking the model back to a lognormal process. To guarantee the existence of a unique pathwise strong solution for~\eqref{main_process>2}, the following existence and uniqueness conditions are required.
\begin{assumption} \label{Assumption3}
    The parameters $-\kappa(t), \theta(t)$, and $-\sigma(t)$ in the NLD-CEV process~\eqref{main_process} are strictly positive and smooth functions depending on the temporal variable $t\in[0,T]$. Moreover, ${\kappa(t)}/{\sigma^{2}(t)}$ is locally bounded on $[0,T]$.
\end{assumption}
\subsection{The NLD-CEV process with regime switching}

Since the NLD-CEV process has limitations in its ability to account for broader economic factors as mentioned in the Introduction, we present a regime-switching version of this model where the~constant parameters are allowed to transition between different states following a Markov chain $X_t \in \mathcal{M}_m$
\begin{equation} \label{model2}
    dR_t  =  \kappa_{X_t}\Big(\theta_{X_t} R_t^{-(1-\beta)}-R_t\Big)dt + \sigma_{X_t} R_t^\frac{\beta}{2}~dW_t^\mathbb{Q},
\end{equation}
where $\mathcal{M}_m := \{1, 2, \ldots, m\}$ denotes the state space. When the situation is believed to be in state $i$, the continuous-time irreducible $m$-state Markov chain with the generator $Q = (q_{ij})_{m \times m}$, which is independent of $W_t^\mathbb{Q}$, is defined as
\begin{align*}
    \mathbb{P}\big(X_{t + \delta} = j \mid X_{t} = i\big) =
    \begin{cases}
        q_{ij}\delta + o(\delta) \ & \text{ if } i \neq j, \\
        1 + q_{ij}\delta + o(\delta) \ & \text{ if } i = j,
    \end{cases}
\end{align*}
where $\delta > 0$ and $o(\delta)$ satisfies $\lim_{\delta\rightarrow0}\frac{o(\delta)}{\delta}=0$. If $i \neq j$, the transition rate from $i$ to $j$ satisfies $q_{ij} \geq 0$ and~$q_{ii} = -\sum_{i\neq j} q_{ij}$. The dynamic behavior of the financial market regime is captured through this Markov chain, which plays a fundamental role in governing the index dynamics. While our analysis focuses on a two-state Markov chain configuration to model regime transitions, the framework can be readily generalized to accommodate any finite number of states. Following the work of Sutthimat, Mekchay and Rujivan~\cite{sutthimat20222closed}, the regime-switching model~\eqref{model2} can be characterized into two distinct, depending  on the parameter $\beta$.

In the case of $\beta\in[0,2)$, we set $\beta = \frac{2\alpha-1}{\alpha}$. For $\alpha\geq\frac{1}{2}$, the regime-switching NLD-CEV process can be written as
\begin{equation} \label{model3}
    dR_t = \kappa_{X_t}\Big(\theta_{X_t}R_t^{\frac{\alpha-1}{\alpha}}-R_t\Big)dt + \sigma_{X_t}R_t^{\frac{2\alpha-1}{2\alpha}} dW_t^\mathbb{Q}.
\end{equation}
In the case of $\beta\in(2,\infty)$, we set $\beta = \frac{2\alpha+1}{\alpha}$. For $\alpha>0$, the regime-switching NLD-CEV process can be written as
\begin{equation} \label{model4}
    dR_t = \kappa_{X_t}\Big(\theta_{X_t}R_t^{\frac{\alpha+1}{\alpha}}-R_t\Big)dt + \sigma_{X_t}R_t^{\frac{2\alpha+1}{2\alpha}}dW_t^\mathbb{Q}.
\end{equation}

This paper proposes an analytical formula for fractional-order conditional moments of the $m$-state regime-switching NLD-CEV process. Specifically, we consider fractional moments of the form
\begin{equation} \label{aim}
   U_{\alpha,i}^{\langle n\rangle}(\tau,R) := \mathbf{E}^{\mathbb{Q}}\Big[ R_T^{\frac{n}{\alpha}} ~|~ R_t= R, X_t = i \Big], \quad 0\leq t \leq T,
\end{equation}
for $\tau = T - t$ denotes the time to maturity, $n \in \mathbb{Z}$, $R > 0$, and the $i$\textsuperscript{th} state is in the state space $\mathcal{M}_m$. The availability of these analytical formulas provides significant value to practitioners in financial markets who require high-precision derivative pricing calculations, particularly when utilizing the $m$-state regime-switching NLD-CEV process to model volatility or interest rate dynamics. A notable contribution came from Ahn and Gao~\cite{ahn1999parametric}, who determined the conditional $\nu^\text{th}$ moments of the $3/2$-SVM in~1999. This model, which aligns with Eq \eqref{main_process} when $\beta = 3$ or Eq \eqref{main_process>2} when $\alpha = 1$, enabled them to analyze their model\rq s distribution using term-structure data. This formulation is also recognized as the IF process, though their derivation relies on integral expressions involving Kummer\rq s and Gamma functions that lack closed-form solutions. Subsequently, Zhou~\cite{zhou2003ito} investigated the conditional moments of the process~\eqref{main_process} with $\beta \in [0, 2)$ in~2003, aiming to implement parameter estimation via the generalized method of moments (GMM). Given the absence of closed-form expressions for the required moments, Zhou developed approximations for the first and second moments via a diffusion process using It\^{o}\rq s lemma. In their 2011 work on the hybrid Heston  model, Grzelak and Oosterlee~\cite{grzelak2011heston} approximated the conditional half-moment of the CIR process (corresponding to the process in~\eqref{main_process<2} with $\alpha = 1$ under constant parameters) via first-order Taylor expansion. Later, Rujivan and Zhu~\cite{rujivan2014simple} derived the first and second conditional moments in~2014 to establish a closed-form solution for discretely sampled variance swaps within the Heston model framework, specifically utilizing the ECIR process.

In the context of conditional expectation, a key question arises: Can we compute the conditional expectation directly from the transition probability density function (PDF)? We begin by introducing the~ECIR process
\begin{equation} \label{eq_ECIR}
    d{V_t} = \kappa(t)(\theta(t) - {V_t})dt + \sigma(t) \sqrt{V_t} \, d{W_t^\mathbb{Q}},
\end{equation}
where the parameter functions; $\theta(t) > 0$, $\kappa(t) > 0$, and $\sigma(t) > 0$ are continuous on $[0, T]$. The transition~PDF of the ECIR process is related to Laguerre polynomials 
$\textbf{L}^{(\eta)}_{k}(x)$ (see also~\cite{mirevski2010some}), the gamma function $\Gamma (x)$, and a time-varying dimension $\textbf{d}(t) := \frac{4\kappa(t)\theta(t)}{\sigma^2(t)}$. Recently, Rujivan et al.~\cite{rujivan2023analytically} presented the transition PDF, $f_{V_t}$, for the ECIR process~\eqref{eq_ECIR}, as follows:
\begin{equation*} \label{defTPDF}
    f_{V_t}(v,t~|~v_0) := \mathbb{P}\left( V_t=v~|~V_0=v_0 \right),
\end{equation*} 
for $v,v_0>0$, and $t\in(0,T]$, which can be expressed as
\begin{equation*} \label{ExxxECIR}
    f_{V_t}(v,t|v_0)=\frac{e^{-\frac{v}{2\tau(t,0)}} v^{\frac{\textbf{d}(t)}{2}-1}}{(2\tau(t,0))^{\frac{\textbf{d}(t)}{2}}} \sum_{k=0}^{\infty} \frac{k!}{\Gamma\big(\frac{\textbf{d}(t)}{2}+k\big)} \hat{c}_{k}(t,v_0) \ \textbf{L}^{\big( \frac{\textbf{d}(t)}{2}-1 \big)}_{k} \Big( \frac{v}{2\tau(t,0)} \Big),
\end{equation*}
where
\begin{align*} \label{c0ECIR}
    \hat{c}_{0}(t,v_0) &= 1, \quad \hat{c}_{k} (t,v_0) = \frac{1}{k} \sum_{j=0}^{k-1} \hat{c}_{j}(t,v_0) \hat{d}_{k-j}(t,v_0), \quad \forall k \in \mathbb{N}, \\
    \hat{d}_{1}(t,v_0) &= -\frac{1}{2\tau(t,0)} v_0 e^{-\int_{0}^{t} \kappa(u) du} +\frac{1}{2} \int_{0}^{t} \textbf{\em d}^{(1)}(s) \left(1-\frac{\tau(t,s)}{\tau(t,0)}\right)ds, \\
    \hat{d}_{j}(t,v_0) &= \frac{1}{2} \int_{0}^{t} \textbf{\em d}^{(1)}(s) \Big(1-\frac{\tau(t,s)}{\tau(t,0)}\Big)^j ds, \quad \forall j \in \mathbb{N}\setminus\{1\},
\end{align*}
and $\tau(t,s) = \frac{1}{4} \int_{s}^{t} \sigma^2(\zeta) e^{-\int_{\zeta}^{t} \kappa(u) du} d\zeta$. In particular, if $\textbf{\em d}(s)=d\geq2$ for all $s\in [0,t]$, then
\begin{equation*} \label{ckCIR}
    \hat{c}_{k} (t,v_0) = \Bigg(-\frac{e^{-\int_{0}^{t} \kappa(u) du} }{2\tau(t,0)} \Bigg)^k \frac{v_{0}^{k}}{k!}, \quad \forall k \in \mathbb{N} \cup \{0\}.
\end{equation*}
Applying It\^{o}\rq s lemma, along with the transformation $V_t = R_t^{2 - \beta}$ in~\eqref{main_process}, yields an ECIR process as~follows
\begin{equation} \label{ECIR}
    dV_t = A(t) \left(B(t) - V_t\right) dt + C(t) \sqrt{V_t} \, dW_t,
\end{equation}
where $A(t) = (2 - \beta) \kappa(t)$, $B(t) = \theta(t) + \frac{(1 - \beta) \sigma^2(t)}{2 \kappa(t)}$, and $C(t) = (2 - \beta) \sigma(t)$. When the parameters $\beta = \frac{2 \alpha - 1}{\alpha}$ and $\beta = \frac{2 \alpha + 1}{\alpha}$ are substituted in~\eqref{ECIR}, they yield the processes~\eqref{main_process<2} and~\eqref{main_process>2}, respectively. This implies that we can directly transform the ECIR process\rq s transition PDF to obtain the transition PDF for the~NLD-CEV process (see also~\cite{ahn1999parametric}).

However, deriving conditional expectations (such as conditional moments) directly from the transition PDF are difficult, and this becomes even more complicated for the conditional moments of the $m$-state regime-switching NLD-CEV process. To overcome this issue, the Feynman--Kac formula for switching diffusions, which is well-suited for SDEs with regime switching, is utilized.
\section{Conditional moments: A hybrid system of PDEs} \label{sec_CM}
\medskip

The results in this section follow a structured progression from general regime-switching dynamics to special and illustrative cases.
We first establish the hybrid coupled PDE systems associated with the $m$-state regime-switching NLD--CEV models~\eqref{model3} and~\eqref{model4} via the switching diffusion Feynman--Kac approach; see Theorems~\ref{thm_hbs1} and~\ref{thm_hbs2}.
These PDE characterizations lead to the general $m$-state finite-sum moment representations~\eqref{Ern} and~\eqref{Ernn}, where the coefficient vectors satisfy the recursive matrix differential equations driven by the Markov generator.
To obtain explicit and computable formulas, the general construction is then specialized to the two-state case under $\beta\in[0,2)$ and $\beta\in(2,\infty)$, resulting in Theorems~\ref{thm11} and~\ref{thm321}, respectively.
Theorems~\ref{thm312} and~\ref{thm7} further consider parameter-homogeneous settings, under which the regime-switching structure becomes degenerate and the resulting moment formulas reduce to those of a non-switching NLD-CEV process.
In addition, Theorems~\ref{thm313} and~\ref{thm323} focus on the first-order fractional moments, providing explicit closed-form expressions that serve as building blocks for long-term moment behavior.
Taking the limit $\tau\to\infty$ of these conditional moments yields the corresponding unconditional moments, which are summarized in Theorems~\ref{thm_uncon_1} and~\ref{thm_uncon_2}, derived, respectively, from Theorems~\ref{thm313} and~\ref{thm323}.

To facilitate the construction of the hybrid PDE systems and ensure the well-posedness of the underlying regime-switching diffusions, we first state the technical assumptions required for the NLD--CEV dynamics.
These assumptions guarantee the existence of a unique pathwise strong solution and allow the application of the switching-diffusion Feynman--Kac formula.
Specifically, Assumptions~\ref{Assumption4} and~\ref{Assumption5} are imposed on Model~\eqref{model3}, while Assumption~\ref{Assumption6} applies to Model~\eqref{model4}.
\begin{assumption} \label{Assumption4}
    For any $X_t \in \mathcal{M}_m$, the parameters $\kappa_{X_t}, \theta_{X_t}$, and $\sigma_{X_t}$ in the NLD-CEV process~\eqref{model3} are strictly positive and smooth functions depending on the temporal variable $t\in[0,T]$. Moreover, ${\kappa_{X_t}}/{\sigma^{2}_{X_t}}$ is locally bounded on $[0,T]$.
\end{assumption}
\begin{assumption} \label{Assumption5}
    For any $X_t \in \mathcal{M}_m$, the process $R_t$ in~\eqref{model3} contains the inequality $2\kappa_{X_t}\theta_{X_t} \geq \sigma_{X_t}^2$.
\end{assumption}
\begin{assumption} \label{Assumption6}
    For any $X_t \in \mathcal{M}_m$, the parameters $-\kappa_{X_t}, \theta_{X_t}$, and $-\sigma_{X_t}$ in the NLD-CEV process~\eqref{model4} are strictly positive and smooth functions depending on the temporal variable $t\in[0,T]$. Moreover, ${\kappa_{X_t}}/{\sigma^{2}_{X_t}}$ is locally bounded on $[0,T]$.
\end{assumption}

In real markets, volatility may evolve through multiple levels rather than a single low--high pattern. Our theoretical derivations, however, do \emph{not} rely on a two-regime restriction: Throughout, the regime process is allowed to be a finite-state continuous-time Markov chain with an arbitrary number of states~$m\ge2$. In particular, the hybrid (coupled) system of PDEs is formulated in full generality in Theorem~\ref{thm_hbs1} (Subsection~\ref{subsec_hybrid_system}) for $\beta\in[0,2)$ and in Theorem~\ref{thm_hbs2} (Subsection~\ref{subsec_hybrid__system}) for $\beta\in(2,\infty)$; moving from $m=2$ to any $m$ simply replaces the $2\times2$ generator with an $m\times m$ generator and yields an $m$-dimensional coupled system of the same structure.

For numerical illustrations, we specialize to the two-state cases in Subsections~\ref{subsec02} and~\ref{subsec2ty}. This choice is motivated by clarity and parsimony: The two-state specification is the simplest nontrivial
regime-switching model that captures basic low--high volatility behavior and keeps the parameter dimension manageable, allowing us to highlight the analytic tractability and computational efficiency of the proposed closed-form pricing formulas. We emphasize that this two-state implementation is used as a transparent baseline example; the same methodology and theoretical results extend directly to multistate regime specifications.

\subsection{Hybrid system of PDEs when {$\beta\in[0,2)$}} \label{subsec_hybrid_system}
\medskip

A hybrid system of PDEs is established from the Feynman--Kac formula for switching diffusions related to the fractional-order conditional moments in~\eqref{aim} and the $m$-state regime-switching NLD-CEV process in~\eqref{model3}, as presented in the following theorem.
\begin{theorem} \label{thm_hbs1}
    Suppose that $R_t$ follows the system of SDEs~\eqref{model3} on $[0,T]$ and Assumptions~\ref{Assumption4} and~\ref{Assumption5} hold. The function $U_{\alpha,i}^{\langle n\rangle} := U_{\alpha,i}^{\langle n\rangle}(\tau,R)$ defined in~\eqref{aim}, for $i \in \mathcal{M}_m$ and $\tau = T - t \geq 0$, are the solutions of the following hybrid system of PDEs:
    \begin{equation} \label{thm_hy1}
          \frac{\partial U_{\alpha,i}^{\langle n\rangle}}{\partial \tau} - \kappa_{i}\Big(\theta_{i}R^{\frac{\alpha-1}{\alpha}}-R\Big) \frac{\partial U_{\alpha,i}^{\langle n\rangle}}{\partial R} - \frac{\sigma_{i}^2 R^{\frac{2\alpha-1}{\alpha}}}{2}\frac{\partial^2 U_{\alpha,i}^{\langle n\rangle}}{\partial R^2} - \sum_{\substack{j \in \mathcal{M}_m \\ j \neq i}} q_{ij}\left( U_{\alpha,j}^{\langle n\rangle} - U_{\alpha,i}^{\langle n\rangle} \right) = 0,
    \end{equation}
    for $(\tau,R) \in [0,T] \times (0,\infty)$ subject to the initial condition \ $U_{\alpha,i}^{\langle n\rangle}(0,R) = R^{\frac{n}{\alpha}}$.
\end{theorem}
\begin{proof}
    The SDE $R_t$ defined in~\eqref{model3} falls within the category of switching diffusion processes studied by Baran et al.~\cite{baran2013feynman}. Its corresponding infinitesimal generator, $\mathcal{L}$, is given by
    \begin{equation} \label{thm_hy1_proof1}
        -\frac{\partial U_{\alpha,i}^{\langle n\rangle}}{\partial t} = \mathcal{L} U_{\alpha,i}^{\langle n\rangle} = \kappa_{i}\Big(\theta_{i}R^{\frac{\alpha-1}{\alpha}}-R\Big) \frac{\partial U_{\alpha,i}^{\langle n\rangle}}{\partial R} + \frac{\sigma_{i}^2 R^{\frac{2\alpha-1}{\alpha}}}{2}\frac{\partial^2 U_{\alpha,i}^{\langle n\rangle}}{\partial R^2} + \sum_{\substack{j \in \mathcal{M}_m \\ j \neq i}} q_{ij}\left( U_{\alpha,j}^{\langle n\rangle} - U_{\alpha,i}^{\langle n\rangle} \right),
    \end{equation}
    and since $\tau = T-t$,~\eqref{thm_hy1_proof1} becomes~\eqref{thm_hy1} as required; see more details in \cite[Theorem 6]{baran2013feynman}.
\end{proof}
Compared with the single-regime NLD-CEV process, the inclusion of a finite-state Markov chain introduces additional analytical difficulties. In particular, regime-switching couples the fractional-order conditional moments across different regimes through the transition intensities $q_{ij}$, leading to a system of hybrid PDEs rather than a single equation. This coupling substantially increases the analytical complexity and prevents a direct extension of the standard moment-based techniques developed for non-switching models.

The solution to~\eqref{thm_hy1} provides closed-form expressions for the fractional-order conditional moments of the process $R_t$. Specifically, these moments can be derived by expressing $\mathbf{E}^{\mathbb{Q}}\big[ R_T^{\frac{n}{\alpha}} \mid R_t= R, X_t = i \big]$ as a sum of the terms involving the rational power of $R$ weighted by the coefficients $A_{\alpha, j}^{\langle k\rangle}(\tau)$, which are determined through a recursive system. This formulation allows us to compute the conditional moments iteratively, capturing the impact of regime-switching dynamics through matrix operations involving $\mathbf{P}_\alpha^{\langle \ell\rangle}$ and $\mathbf{D}_\alpha^{\langle \ell\rangle}$. By solving for the coefficients $A_{\alpha, j}^{\langle \ell \rangle}(\tau)$ at each step, the fractional-order conditional moments of $R_t$ can be expressed as follows:
\begin{equation} \label{Ern}
    \mathbf{E}^{\mathbb{Q}}\Big[ R_T^{\frac{n}{\alpha}} ~|~ R_t= R, X_t = i \Big] = \sum_{j = 1}^{m} \Bigg( \mathds{1}_{ \{j\} }(i)~ \sum_{k=0}^n A_{\alpha, j}^{\langle k\rangle}(\tau) R^\frac{k}{\alpha} \Bigg),
\end{equation}
where $\tau = T-t\geq0$ and $\mathds{1}$ is the indicator function. The coefficients $A_{\alpha, j}^{\langle n \rangle}(\tau)$ for $j = 1, 2, \ldots, m$ \, are the solutions of
\begin{equation*}
    \mathbf{\overline{u}}_\alpha^{\langle n\rangle}(\tau) = \mathbf{P}_\alpha^{\langle n\rangle} \mathbf{u}_\alpha^{\langle n\rangle}(\tau).
\end{equation*}
For $\ell=n-1, n-2, \dots, 0$, $A_{\alpha, j}^{\langle \ell \rangle}(\tau)$ for $j = 1, 2, \ldots, m$, can be solved backward iteratively through the following system:
\begin{equation*}
    \mathbf{\overline{u}}_\alpha^{\langle \ell\rangle}(\tau) = \mathbf{P}_\alpha^{\langle \ell\rangle} \mathbf{u}_\alpha^{\langle \ell\rangle}(\tau) + \mathbf{D}_\alpha^{\langle \ell\rangle} \mathbf{u}_\alpha^{\langle \ell+1\rangle}(\tau),
\end{equation*}
where \, $\mathbf{u}_\alpha^{\langle \imath \rangle}(\tau) = \big[ A_{\alpha, 1}^{\langle \imath \rangle}(\tau), A_{\alpha, 2}^{\langle \imath \rangle}(\tau), \ldots, A_{\alpha, m}^{\langle \imath \rangle}(\tau) \big]^\top$, \, $\mathbf{\overline{u}}_\alpha^{\langle \imath \rangle}(\tau) = \big[ \frac{d}{d\tau} A_{\alpha, 1}^{\langle \imath \rangle}(\tau), \frac{d}{d\tau} A_{\alpha, 2}^{\langle \imath \rangle}(\tau), \ldots, \frac{d}{d\tau} A_{\alpha, m}^{\langle \imath \rangle}(\tau) \big]^\top$, for~$0 \leq \imath \leq n$,
\begin{gather*}
    \mathbf{P}_\alpha^{\langle \ell\rangle} =
    \begin{bmatrix}
        -\left( \kappa_1\frac{\ell}{\alpha} + \sum_{\imath=2}^m q_{1,\imath} \right) & q_{1,2} & \ldots & q_{1,m-1}& q_{1,m} \\ q_{2,1} & -\left( \kappa_2\frac{\ell}{\alpha} + {\sum_{\imath=1, \imath \neq 2}^m} q_{2,\imath} \right) & \ldots &q_{2,m-1}& q_{2,m} \\  \vdots & \vdots & \ddots & \vdots & \vdots \\ q_{m,1} & q_{m,2} & \ldots & q_{m,m-1} & -\left( \kappa_m\frac{\ell}{\alpha} + \sum_{\imath=1}^{m-1} q_{m,\imath} \right)
    \end{bmatrix},
\end{gather*}
and \ $\mathbf{D}_\alpha^{\langle \ell\rangle} = \mathrm{diag}\big\{ \gamma_{\alpha,1}^{\langle\ell\rangle}, \gamma_{\alpha,2}^{\langle\ell\rangle}, \ldots, \gamma_{\alpha,m}^{\langle\ell\rangle} \big\}$ \ when \ $\gamma_{\alpha,i}^{\langle\ell\rangle} = \left( \frac{\ell+1}{\alpha} \right)\left(\kappa_i\theta_i+\frac{1}{2}\sigma_i^2 \left( \frac{\ell+1}{\alpha}-1\right) \right)$; see a rigorous proof for the two-state regime-switching NLD-CEV process in Section~\ref{subsec02}.

For clarity, we briefly summarize the structure and interpretation of the recursive system above before specializing to the two-state case.
The recursive coefficient system admits a clear structural interpretation.
At the highest order $k=n$, the coefficients satisfy a homogeneous linear system governed by the mean-reversion parameters and the Markov transition intensities.
Lower-order coefficients are then obtained sequentially by solving linear inhomogeneous systems, where the source terms depend explicitly on the coefficients of higher order.
This backward recursive structure reflects the polynomial expansion of the conditional moments and guarantees closure of the moment system in a finite number of steps.
Moreover, the regime-switching effects are propagated across different moment orders through the coupling matrices $\mathbf{P}_\alpha^{\langle \ell\rangle}$ and $\mathbf{D}_\alpha^{\langle \ell\rangle}$, providing a transparent mechanism for incorporating regime dynamics into the moment's evolution.

\subsection{Hybrid system of PDEs when $\beta\in(2,\infty)$}\label{subsec_hybrid__system}
\medskip

A hybrid system of PDEs is established from the Feynman--Kac formula for switching diffusions related to the fractional-order conditional moments in~\eqref{aim} and the $m$-state regime-switching NLD-CEV process in~\eqref{model4}, as presented in the following theorem.
\begin{theorem} \label{thm_hbs2}
    Suppose that $R_t$ follows the system of SDEs~\eqref{model4} on $[0,T]$ and that Assumption~\ref{Assumption6} holds. The function $U_{\alpha,i}^{\langle n\rangle} := U_{\alpha,i}^{\langle n\rangle}(\tau,R)$ defined in~\eqref{aim}, for $i \in \mathcal{M}_m$ and $\tau = T - t \geq 0$, are the solutions of the following hybrid system of PDEs:
    \begin{equation} \label{thm_hy2}
          \frac{\partial U_{\alpha,i}^{\langle n\rangle}}{\partial \tau} - \kappa_{i}\Big(\theta_{i}R^{\frac{\alpha+1}{\alpha}}-R\Big) \frac{\partial U_{\alpha,i}^{\langle n\rangle}}{\partial R} - \frac{\sigma_{i}^2 R^{\frac{2\alpha+1}{\alpha}}}{2}\frac{\partial^2 U_{\alpha,i}^{\langle n\rangle}}{\partial R^2} - \sum_{\substack{j \in \mathcal{M}_m \\ j \neq i}} q_{ij}\left( U_{\alpha,j}^{\langle n\rangle} - U_{\alpha,i}^{\langle n\rangle} \right) = 0,
    \end{equation}
    for $(\tau,R) \in [0,T] \times (0,\infty)$ subject to the initial condition \ $U_{\alpha,i}^{\langle n\rangle}(0,R) = R^{\frac{n}{\alpha}}$.
\end{theorem}
\begin{proof}
    See Theorem~6 in Baran et al.~\cite{baran2013feynman} and Theorem~\ref{thm_hbs1}.
\end{proof}
As described previously in Section~\ref{subsec_hybrid_system}, solving for the coefficients $A_{\alpha, j}^{\langle \ell \rangle}(\tau)$ at each step, the fractional-order conditional moments of $R_t$, which is the solution of the system of PDEs in~\eqref{thm_hy2}, can be expressed as follows:
\begin{equation} \label{Ernn}
    \mathbf{E}^{\mathbb{Q}}\Big[ R_T^{\frac{n}{\alpha}} ~|~ R_t= R, X_t = i \Big] = \sum_{j = 1}^{m} \Bigg( \mathds{1}_{ \{j\} }(i)~ \sum_{k=0}^{|n|} A_{\alpha, j}^{\langle k\rangle}(\tau) R^{-\frac{k}{\alpha}} \Bigg),
\end{equation}
where $\tau = T-t\geq0$ and $\mathds{1}$ is the indicator function. The coefficients $A_{\alpha, j}^{\langle n \rangle}(\tau)$ for $j = 1, 2, \ldots, m$ \, are the solutions of
\begin{equation*}
    \mathbf{\overline{u}}_\alpha^{\langle |n| \rangle}(\tau) = \mathbf{P}_\alpha^{\langle |n| \rangle} \mathbf{u}_\alpha^{\langle |n| \rangle}(\tau),
\end{equation*}
and $A_{\alpha, j}^{\langle \ell \rangle}(\tau)$ for $j = 1, 2, \ldots, m$, can be solved iteratively via the following system:
\begin{equation*}
    \mathbf{\overline{u}}_\alpha^{\langle \ell\rangle}(\tau) = \mathbf{P}_\alpha^{\langle \ell\rangle} \mathbf{u}_\alpha^{\langle \ell\rangle}(\tau) + \mathbf{D}_\alpha^{\langle \ell\rangle} \mathbf{u}_\alpha^{\langle \ell+1\rangle}(\tau), \quad 0 \leq \ell < |n|,
\end{equation*}
where
\begin{gather*}
    \mathbf{P}_\alpha^{\langle \ell\rangle} =
    \begin{bmatrix}
        \kappa_1\frac{\ell}{\alpha} - \sum_{\imath=2}^m q_{1,\imath} & q_{1,2} & \ldots & q_{1,m-1}& q_{1,m} \\ q_{2,1} &  \kappa_2\frac{\ell}{\alpha} - {\sum_{\imath=1, \imath \neq 2}^m} q_{2,\imath} & \ldots &q_{2,m-1}& q_{2,m} \\  \vdots & \vdots & \ddots & \vdots & \vdots \\ q_{m,1} & q_{m,2} & \ldots & q_{m,m-1} & \kappa_m\frac{\ell}{\alpha} - \sum_{\imath=1}^{m-1} q_{m,\imath}
    \end{bmatrix},
\end{gather*}
and \ $\mathbf{D}_\alpha^{\langle \ell\rangle} = \mathrm{diag}\big\{ \overline\gamma_{\alpha,1}^{\langle\ell\rangle}, \overline\gamma_{\alpha,2}^{\langle\ell\rangle}, \ldots, \overline\gamma_{\alpha,m}^{\langle\ell\rangle} \big\}$ \ when \ $\overline\gamma_{\alpha,i}^{\langle\ell\rangle} = \left( \frac{\ell+1}{\alpha} \right)\left( - \kappa_i\theta_i+\frac{1}{2}\sigma_i^2 \left( \frac{\ell+1}{\alpha} + 1\right) \right)$; see a rigorous proof for the two-state regime-switching NLD-CEV process in Section~\ref{subsec2ty}.

A key feature of the proposed analytical formulas~\eqref{Ern} and~\eqref{Ernn} is the explicit role of the $m$-state regime-switching intensities $q_{ij}$, which couple the fractional-order conditional moments across different regimes via terms of the form
$q_{ij}\big(U_{\alpha,j}^{\langle n\rangle} - U_{\alpha,i}^{\langle n\rangle}\big)$ for $i \neq j$
in the hybrid PDE systems~\eqref{thm_hy1} and~\eqref{thm_hy2}. When the transition intensities are relatively small, the regime-switching time scale is slow compared with the diffusion dynamics, leading to weak coupling between regimes and conditional moments that evolve close to their single-regime counterparts. In contrast, when the transition intensities are relatively large, the regime-switching time scale becomes short, resulting in strong coupling across regimes and conditional moments that tend to exhibit an averaged behavior influenced by the stationary distribution of the underlying Markov chain. This qualitative dependence of conditional moments on the relative magnitude of regime-switching intensities enables the proposed analytical formulation to accommodate a wide range of market conditions, from persistent regimes to environments characterized by frequent regime changes.

\subsection{The two-state regime-switching NLD-CEV process when {$\beta\in[0,2)$}} \label{subsec02}
\medskip

This section focuses on deriving an explicit formula for the fractional-order conditional moments of the two-state regime-switching NLD-CEV process under the parameter range $\beta \in [0, 2)$. Building on the hybrid system approach introduced earlier, the conditional moments are derived by applying a system of PDEs tailored to the NLD-CEV dynamics under regime switching. The following theorem establishes the formal solution to this system, offering insight into the behavior of conditional moments within a two-state switching process. Before proceeding with the theorem, and following statements in Section~\ref{subsec_hybrid_system}, let $A_\alpha^{\langle k\rangle}(\tau) := A_{\alpha, 1}^{\langle k\rangle}(\tau)$ and $B_\alpha^{\langle k\rangle}(\tau) := A_{\alpha, 2}^{\langle k\rangle}(\tau)$.
\begin{theorem} \label{thm11}
    Suppose that $R_t$ follows the system of SDEs~\eqref{model3} on $[0,T]$ with the initial conditions $R_{t}=R$ and $X_{t}=i,i\in\{1,2\}$. The conditional $\frac{n}{\alpha}$-moment for $n\in\mathbb{Z}^+_0$ is
    \begin{equation} \label{cm1}
        U_{\alpha,i}^{\langle n\rangle}(\tau,R)
        = \mathbf{E}^{\mathbb{Q}}\Big[ R_T^{\frac{n}{\alpha}} ~|~ R_t= R, X_t = i \Big] 
        = \mathds{1}_{ \{1\} }(i)~ \sum_{k=0}^n A_\alpha^{\langle k\rangle}(\tau)R^\frac{k}{\alpha} + \mathds{1}_{ \{2\} }(i)~  \sum_{k=0}^n B_\alpha^{\langle k\rangle}(\tau)R^\frac{k}{\alpha},
    \end{equation}
    where $\tau = T-t\geq0$, $\mathds{1}$ is the indicator function, and $A_\alpha^{\langle k\rangle}$ and $B_\alpha^{\langle k\rangle}$ for $k=0,1,\ldots,n$ can be solved by the system of recursive matrix differential equations
    \begin{align}
        \mathbf{\overline{u}}_\alpha^{\langle n\rangle}(\tau)
        &= \mathbf{P}_\alpha^{\langle n\rangle}
        \mathbf{u}_\alpha^{\langle n\rangle}(\tau) \label{311}, \\
        \mathbf{\overline{u}}_\alpha^{\langle \ell\rangle}(\tau)
        &= \mathbf{P}_\alpha^{\langle \ell\rangle}
        \mathbf{u}_\alpha^{\langle \ell\rangle}(\tau)
        + \mathbf{D}_\alpha^{\langle \ell\rangle}
        \mathbf{u}_\alpha^{\langle \ell+1\rangle}(\tau), \label{312} 
    \end{align}
    with the initial conditions $\mathbf{u}_\alpha^{\langle n\rangle}(0)=[1,1]^\top$ and $\mathbf{u}_\alpha^{\langle \ell\rangle}(0)=[0,0]^\top~\mbox{for}~\ell=0,1,\ldots,n-1$, where
    \begin{align*}
        &\mathbf{\overline{u}}_\alpha^{\langle k\rangle}(\tau) =
        \begin{bmatrix}
            \frac{d}{d\tau}A_\alpha^{\langle k\rangle}(\tau) \\ \frac{d}{d\tau}B_\alpha^{\langle k\rangle}(\tau)
            \end{bmatrix}, \quad
        \mathbf{P}_\alpha^{\langle k\rangle} =
        \begin{bmatrix}
            -\kappa_1\frac{k}{\alpha}-q_{12} &q_{12} \\q_{21}&-\kappa_2\frac{k}{\alpha}-q_{21}
        \end{bmatrix}, \quad
        \mathbf{u}_\alpha^{\langle k\rangle}(\tau) =
        \begin{bmatrix}
            A_\alpha^{\langle k\rangle}(\tau)\\B_\alpha^{\langle k\rangle}(\tau)
        \end{bmatrix}, \\
        &\mathbf{D}_\alpha^{\langle \ell\rangle} = \mathrm{diag}\left\{\gamma_{\alpha,1}^{\langle\ell\rangle},\gamma_{\alpha,2}^{\langle\ell\rangle}\right\}, \quad \gamma_{\alpha,i}^{\langle\ell\rangle} =
        \left(\frac{\ell+1}{\alpha}\right)\left(\kappa_i\theta_i+\frac{1}{2}\sigma_i^2\left(\frac{\ell+1}{\alpha}-1\right)\right).
    \end{align*}
\end{theorem}
\begin{proof}
    By the result of Yao et al.~\cite{yao2006regime}, the conditional moments in~\eqref{cm1} satisfy the PDE
    \begin{equation} \label{Yao.2006}
        \frac{\partial U_{\alpha,i}^{\langle n\rangle}}{\partial \tau}-
        \kappa_i\left(\theta_iR^{-(1-\beta)}-R\right)\frac{\partial U_{\alpha,i}^{\langle n\rangle}}{\partial R}
        - \frac{1}{2}\sigma_i^2R^{\beta}\frac{\partial^2 U_{\alpha,i}^{\langle n\rangle}}{\partial R^2}
        -\sum_{j \neq i} q_{ij}\left(U_{\alpha,j}^{\langle n\rangle} - U_{\alpha,i}^{\langle n\rangle}\right) = 0,
     \end{equation}
    with $\beta = \frac{2\alpha-1}{\alpha}$ where $(\tau,R) \in [0,T] \times \mathbb{R^+}$ subject to the initial condition at $\tau=0$
    \begin{align*}
        U_{\alpha,i}^{\langle n\rangle}(0,R) = R^\frac{n}{\alpha} \quad \mbox{for}~i = 1, 2.
    \end{align*}
    The proof is divided into two cases depending on the state $X_t = i$ for $i = 1, 2$. For $i = 1$, we obtain the initial conditions $A_\alpha^{\langle n\rangle}(0) = 1$ by comparing the coefficients of~\eqref{cm1}. Next, we compute  Eq \eqref{Yao.2006} by substituting its partial derivatives with respect to~\eqref{cm1}, which are
    \begin{equation*}
        \frac{\partial U_{\alpha,i}^{\langle n\rangle}}{\partial \tau} =
        \sum_{k=0}^n\frac{d}{d\tau}A_\alpha^{\langle k\rangle}(\tau)R^\frac{k}{\alpha},\quad
        \frac{\partial U_{\alpha,i}^{\langle n\rangle}}{\partial R} =
        \sum_{k=1}^n\frac{k}{\alpha}A_\alpha^{\langle k\rangle}(\tau)R^{\frac{k}{\alpha}-1} \quad \mbox{\text{and}} \quad
        \frac{\partial^2 U_{\alpha,i}^{\langle n\rangle}}{\partial R^2} =
        \sum_{k=1}^n\frac{k}{\alpha}\left(\frac{k}{\alpha}-1\right)A_\alpha^{\langle k\rangle}(\tau)R^{\frac{k}{\alpha}-2},
    \end{equation*}
    to obtain the simplified form
    \begin{align}
        &\left(\frac{d}{d\tau}A_\alpha^{\langle n\rangle}(\tau)
            + \kappa_1\frac{n}{\alpha}A_\alpha^{\langle n\rangle}(\tau)
            -q_{12}\left( B_\alpha^{\langle n\rangle}(\tau)
            -A_\alpha^{\langle n\rangle}(\tau)\right)\right) R^\frac{n}{\alpha} \notag \\
            &\quad+\sum_{\ell=0}^{n-1}\Bigg(\frac{d}{d\tau}A_\alpha^{\langle \ell\rangle}(\tau)
            -\kappa_1\theta_1\left(\frac{\ell+1}{\alpha}\right)A_\alpha^{\langle \ell+1\rangle}(\tau)
            +\kappa_1\frac{\ell}{\alpha}A_\alpha^{\langle \ell\rangle}(\tau)
            -\frac{1}{2}\sigma_1^2\left(\frac{\ell+1}{\alpha}\right)\left(\frac{\ell+1}{\alpha}-1\right)A_\alpha^{\langle \ell+1\rangle}(\tau) \notag \\
            &\quad\quad -q_{12} \left(B_\alpha^{\langle \ell\rangle}(\tau)
            -A_\alpha^{\langle \ell\rangle}(\tau)\right)\Bigg)R^\frac{\ell}{\alpha}
            = 0, \label{313}
    \end{align}
    with the initial conditions $A_\alpha^{\langle \ell\rangle}(0)=0$ for all $\ell = 0, 1, \ldots, n-1$. Consequently, we obtain the case of the state~$i = 2$ by directly following the previous case
    \begin{align}
        &\left(\frac{d}{d\tau}B_\alpha^{\langle n\rangle}(\tau)
        +\kappa_2\frac{n}{\alpha}B_\alpha^{\langle n\rangle}(\tau)
        -q_{21}\left( A_\alpha^{\langle n\rangle}(\tau)
        -B_\alpha^{\langle n\rangle}(\tau)\right)\right)R^\frac{n}{\alpha} \notag \\
        &\quad+\sum_{\ell=0}^{n-1}\Bigg(\frac{d}{d\tau}B_\alpha^{\langle \ell\rangle}(\tau)
        -\kappa_2\theta_2\left(\frac{\ell+1}{\alpha}\right)B_\alpha^{\langle \ell+1\rangle}(\tau)
        +\kappa_2\frac{\ell}{\alpha}B_\alpha^{\langle \ell\rangle}(\tau)
        -\frac{1}{2}\sigma_2^2\left(\frac{\ell+1}{\alpha}\right)\left(\frac{\ell+1}{\alpha}-1\right)B_\alpha^{\langle \ell+1\rangle}(\tau) \notag \\
        &\quad\quad -q_{21} \left(A_\alpha^{\langle \ell\rangle}(\tau)
        -B_\alpha^{\langle \ell\rangle}(\tau)\right)\Bigg)R^\frac{\ell}{\alpha} = 0, \label{314}
    \end{align}
    where $B_\alpha^{\langle n\rangle}(0)=1$ and $B_\alpha^{\langle \ell\rangle}(0)=0$ for all $\ell=0,1,\ldots,n-1$. As~\eqref{313} and~\eqref{314} are equivalent to~\eqref{311} and~\eqref{312}, the proof is complete.
\end{proof}

The following theorem addresses the case where the parameters of the both states are similar, allowing for a simplified analysis of the regime-switching NLD-CEV process. This condition enables the derivation of more tractable expressions for the conditional moments, as the similarities in the parameters reduce the complexity of the resulting system of equations. By focusing on this specific scenario, the theorem provides insights into the process' behavior under near-identical regime~conditions.
\begin{theorem} \label{thm312}
    Suppose that $R_t$ follows the system of SDEs~\eqref{model3} such that $\kappa_1=\kappa_2=\kappa, \theta_1=\theta_2=\theta$, and $\sigma_1=\sigma_2=\sigma$ on $[0,T]$ with the initial conditions $R_{t}=R$ and $X_{t}=i,i\in\{1,2\}$. The conditional $\frac{n}{\alpha}$-moment for $n\in\mathbb{Z}^+_0$ is defined by
    \begin{equation} \label{3211}
        \mathbf{E}^{\mathbb{Q}}\Big[ R_T^{\frac{n}{\alpha}} ~|~ R_t = R, X_t = i \Big]
        = \sum_{k=0}^n \left( \frac{e^{-\frac{n\kappa\tau}{\alpha }}}{(n-k)!}\left( \frac{\alpha e^{\frac{\kappa\tau}{\alpha}}-\alpha}{\kappa} \right)^{n-k} \left( \prod_{j=1}^{n-k} \gamma_\alpha^{\langle n-j \rangle} \right) \right) R^\frac{k}{\alpha},
    \end{equation}
    where the product\rq s term equals $1$ when $k = n$.
\end{theorem}
\begin{proof}
    We use the eigenvalue method to solve the system of ordinary differential equations (ODEs) with constant coefficients. The eigenvalues and eigenvectors of $\mathbf{P}_\alpha^{\langle k\rangle}$ are, respectively, $\lambda_1^{\langle k \rangle} = -\kappa\frac{k}{\alpha}$ and~$\lambda_2^{\langle k \rangle} = -q_{12}-q_{21}-\kappa\frac{k}{\alpha}$, corresponding to $\mathbf{v}_1=[1,1]^\top$ and $\mathbf{v}_2 = [q_{12},-q_{21}]^\top$ for $k=0,1,\ldots,n$. Let $S = [\mathbf{v_1}, \mathbf{v_2}]$ 
    and $\Lambda_\alpha^{\langle k\rangle}=\mathrm{diag}\big\{\lambda_1^{\langle k \rangle}, \lambda_2^{\langle k \rangle}\big\}$. From~\cite{wood2004chain}, the solution of the homogeneous problem~\eqref{311} can be solved as~follows:
    \begin{equation} \label{315}
        \mathbf{u}_\alpha^{\langle n\rangle}(\tau)
        = e^{\tau \mathbf{P}_\alpha^{\langle n\rangle}}\mathbf{u}_\alpha^{\langle n\rangle}(0)
        = S e^{\tau\Lambda_\alpha^{\langle n\rangle}}S^{-1}\mathbf{u}_\alpha^{\langle n\rangle}(0)
        = \left[e^{-\frac{\kappa n \tau}{\alpha }}, e^{-\frac{\kappa n \tau}{\alpha }}\right]^\top.
    \end{equation}
    The solution to the nonhomogeneous problem~\eqref{312} is a combination of the solution to the homogeneous Eq \eqref{311}, $\mathbf{u}_{\alpha,c}^{\langle \ell\rangle}$, and a particular solution, $\mathbf{u}_{\alpha,p}^{\langle \ell\rangle}$
    \begin{equation} \label{323}
        \mathbf{u}_\alpha^{\langle \ell\rangle}(\tau)
        = \mathbf{u}_{\alpha,c}^{\langle \ell\rangle}(\tau)+\mathbf{u}_{\alpha,p}^{\langle \ell\rangle}(\tau).
    \end{equation}
    Since the solution of the homogeneous part with respect to $\mathbf{u}_{\alpha,c}^{\langle \ell\rangle}(0)=[0,0]^\top~\mbox{for}~\ell=0,1,\ldots,n-1$, we use the same idea as~\eqref{315} to obtain $\mathbf{u}_{\alpha,c}^{\langle \ell\rangle}(\tau)=[0,0]^\top$. In this part of the solution, we assume~that
    \begin{equation} \label{324}
        \mathbf{u}_{\alpha,p}^{\langle \ell\rangle}(\tau)
        = \Phi_\alpha^{\langle \ell\rangle}(\tau)\mathbf{f}(\tau),
    \end{equation}
    where $\Phi_\alpha^{\langle \ell\rangle}(\tau) = Se^{\tau\Lambda_\alpha^{\langle \ell\rangle}}$ and $\mathbf{f}(\tau)$ is the unknown
    parameter function. From~\eqref{312},
    \begin{align*}
        \frac{d}{d\tau}\mathbf{u}_{\alpha,p}^{\langle \ell\rangle}(\tau)
        &= \frac{d}{d\tau}\Phi_\alpha^{\langle \ell\rangle}(\tau)\mathbf{f}(\tau)+
        \Phi_\alpha^{\langle \ell\rangle}(\tau)\frac{d}{d\tau}\mathbf{f}(\tau) \\
        &= \mathbf{P}_\alpha^{\langle \ell\rangle}\Phi_\alpha^{\langle \ell\rangle}(\tau)\mathbf{f}(\tau)+
        \Phi_\alpha^{\langle \ell\rangle}(\tau)\frac{d}{d\tau}\mathbf{f}(\tau) \\
        &= \mathbf{P}_\alpha^{\langle \ell\rangle}\mathbf{u}_{\alpha,p}^{\langle \ell\rangle}(\tau) + \Phi_\alpha^{\langle \ell\rangle}(\tau)\frac{d}{d\tau}\mathbf{f}(\tau),
    \end{align*}
    which implies that
    \begin{align*}
        \mathbf{P}_\alpha^{\langle \ell\rangle}
        \mathbf{u}_\alpha^{\langle \ell\rangle}(\tau)
        + \mathbf{D}_\alpha^{\langle \ell\rangle}
        \mathbf{u}_\alpha^{\langle \ell+1\rangle}(\tau)
        = \mathbf{P}_\alpha^{\langle \ell\rangle}\mathbf{u}_{\alpha,p}^{\langle \ell\rangle}(\tau) + \Phi_\alpha^{\langle \ell\rangle}(\tau)\frac{d}{d\tau}\mathbf{f}(\tau).
    \end{align*}
    The unknown parameter function $\mathbf{f}(\tau)$ can then be written in the definite integral form as
    \begin{align}
        \mathbf{f}(\tau) = \int_0^\tau\left(\Phi_\alpha^{\langle \ell\rangle}(\xi)\right)^{-1}
        \mathbf{D}^{\langle \ell \rangle}_\alpha\mathbf{u}_{\alpha}^{\langle \ell + 1\rangle}(\xi)d\xi. \label{325}
    \end{align}
    We conclude from~\eqref{323}--\eqref{325} that the solution of the nonhomogeneous problem~\eqref{312} is represented by
    \begin{align}
        \mathbf{u}_\alpha^{\langle \ell\rangle}(\tau)
        &= \Phi_\alpha^{\langle \ell\rangle}(\tau)
        \int_0^\tau\left(\Phi_\alpha^{\langle \ell\rangle}(\xi)\right)^{-1} \mathbf{D}^{\langle \ell \rangle}_\alpha \mathbf{u}_{\alpha}^{\langle \ell+1\rangle}(\xi)d\xi. \label{326}
    \end{align}
    The following benefit formulas are provided to solve~\eqref{326} based on the specific index $\ell$ for each case:
    \begin{align*}
        \Phi_\alpha^{\langle \ell\rangle}(\tau)
        \begin{bmatrix}
            x \\
            0
        \end{bmatrix}
        &= Se^{\tau\Lambda_\alpha^{\langle \ell\rangle}}
        \begin{bmatrix}
            x \\
            0
        \end{bmatrix}
        =
        \begin{bmatrix}
            x e^{-\tau\kappa\frac{\ell}{\alpha }} \\
            x e^{-\tau\kappa\frac{\ell}{\alpha }}
        \end{bmatrix} \quad \mbox{\text{and}} \quad
        \left( \Phi_\alpha^{\langle \ell\rangle}(\tau)\right)^{-1} 
        \begin{bmatrix}
            y \\
            y
        \end{bmatrix}
        = e^{-\tau\Lambda_\alpha^{\langle \ell\rangle}}S^{-1}
        \begin{bmatrix}
            y \\
            y
        \end{bmatrix}
        =\begin{bmatrix}
            y e^{\tau\kappa\frac{\ell}{\alpha }} \\
            0
        \end{bmatrix},
    \end{align*}
    where $x$ and $y$ are arbitrary functions, which yield inductively to capture
    \begin{equation} \label{327}
        \mathbf{u}_\alpha^{\langle \ell\rangle}(\tau)
        = \left(\prod_{j=1}^{n-\ell} \mathbf{D}^{\langle n-j \rangle}_\alpha \right)
        e^{-\tau\kappa\frac{\ell}{\alpha }}
        \begin{bmatrix}
            J_{n-\ell}(\tau)\\
            J_{n-\ell}(\tau)    
        \end{bmatrix},
    \end{equation}
    for all $\ell = 0, 1, \dots, n-1$, where 
    \begin{align}
        J_{k}(\tau) = 
        \int_0^\tau e^{-\frac{\kappa\xi_k}{\alpha }}
        \int_0^{\xi_{k}} e^{-\frac{\kappa\xi_{k-1}}{\alpha }}
        \ \dots
        \int_0^{\xi_{3}} e^{-\frac{\kappa\xi_2}{\alpha }}
        \int_0^{\xi_{2}} e^{-\frac{\kappa\xi_1}{\alpha }}
        d\xi_1 d\xi_2 \dots d\xi_{k-1} d\xi_{k}.\label{328}
    \end{align}
    The definite repeated integrals~\eqref{328} are rewritten by substituting $F(\xi_i)=\displaystyle\int e^{-\frac{\kappa\xi_i}{\alpha}}d\xi_i$ to obtain
    \begin{align}
        J_{k}(\tau)
        &= \int_{F(0)}^{F(\tau)} 
        \int_{F(0)}^{F(\xi_k)} 
        \dots
        \int_{F(0)}^{F(\xi_3)} 
        \int_{F(0)}^{F(\xi_2)} ~ 1 ~
        \, dF(\xi_1)~dF(\xi_2) \dots dF(\xi_{k-1})~dF(\xi_k)\notag\\
        &= \frac{1}{k!}\left(F(\tau)-F(0)\right)^k\notag\\
        &= \frac{1}{k!}\left(\frac{-\alpha e^{-\frac{\kappa\tau}{\alpha}}+\alpha}{\kappa}\right)^k, \label{329}
    \end{align}
    where we use \cite[Proposition 2.3]{haddad2021repeated} in the second equality. Since both components of the solution~\eqref{327} are the same, we have
    \begin{equation} \label{3210}
        A^{\langle\ell\rangle}_\alpha(\tau) = B^{\langle\ell\rangle}_\alpha(\tau)
        = \frac{e^{\frac{-n\kappa\tau}{\alpha }}}{(n-\ell)!}\left(\frac{\alpha e^{\frac{\kappa\tau}{\alpha}}-\alpha}{\kappa}\right)^{n-\ell}
        \left(\prod_{j=1}^{n-\ell} \gamma^{\langle n-j \rangle}_\alpha \right),
    \end{equation}
    for all $\ell=0,1,\dots,n-1$. When $\ell=n$, Eq \eqref{3210} with the product term set equal to $1$ is equivalent to~\eqref{315}. By~\eqref{cm1} and~\eqref{3210}, the proof is complete.
\end{proof}
It is worth noting that Formula~\eqref{3211} corresponds directly to \cite[(17)]{sutthimat20222closed}, highlighting the consistency between these results. The following theorem also illustrates that Theorem~\ref{thm11} can be effectively applied to calculate the conditional $\frac{1}{\alpha}$-moment, providing a practical method for evaluating fractional-order moments within the NLD-CEV process under specific conditions.
\begin{theorem} \label{thm313}
    Suppose that $R_t$ follows the system of SDEs~\eqref{model3} on $[0,T]$ with the initial conditions $R_{t}=R$ and $X_{t}=i,i \in \{1,2\}$. The conditional $\frac{1}{\alpha}$-moment is defined by
    \begin{equation*}
        \mathbf{E}^{\mathbb{Q}} \Big[ R_T^{\frac{1}{\alpha}} ~|~ R_t= R, X_t = i \Big] 
        = \mathds{1}_{ \{1\} }(i)
        \left(A_\alpha^{\langle 0\rangle}(\tau)+A_\alpha^{\langle 1\rangle}(\tau)R^\frac{1}{\alpha}\right)
        +\mathds{1}_{ \{2\} }(i)
        \left(B_\alpha^{\langle 0\rangle}(\tau)+B_\alpha^{\langle 1\rangle}(\tau)R^\frac{1}{\alpha}\right),
    \end{equation*}
    where $\tau = T-t \geq 0$, $\mathds{1}$ is the indicator function, and $A_\alpha^{\langle k\rangle}$ and $B_\alpha^{\langle k\rangle}$ for $k=0,1$ can be calculated by the~following
    \begin{align*}
        A_\alpha^{\langle 1\rangle}(\tau)
        &= m_{11}e^{\tau\lambda_1^{\langle 1 \rangle}} + m_{12}e^{\tau\lambda_2^{\langle 1 \rangle}}, \quad B_\alpha^{\langle 1\rangle}(\tau)
        = m_{21}e^{\tau\lambda_1^{\langle 1 \rangle}} + m_{22}e^{\tau\lambda_2^{\langle 1 \rangle}}, \\
        A_\alpha^{\langle 0\rangle}(\tau)
        &= \frac{n_{11} \big( e^{\tau\lambda_1^{\langle 1 \rangle} }-1 \big)}{\lambda_1^{\langle 1 \rangle}}+\frac{n_{12} \big( e^{\tau\lambda_2^{\langle 1 \rangle} }-1 \big)}{\lambda_2^{\langle 1 \rangle}} + \frac{n_{21} q_{12} \big( e^{\tau\lambda _1^{\langle 1 \rangle} }-e^{-\tau(q_{12} + q_{21}) } \big)}{\lambda_1^{\langle 1 \rangle}+q_{12}+q_{21} } + \frac{n_{22} q_{12} \big( e^{\tau\lambda _2^{\langle 1 \rangle} }-e^{-\tau(q_{12}+q_{21})} \big)}{\lambda_2^{\langle 1 \rangle} +q_{12}+q_{21} }, \\
        B_\alpha^{\langle 0\rangle}(\tau)
        &= \frac{n_{11} \big(e^{\tau \lambda _1^{\langle 1 \rangle} }-1\big)}{\lambda _1^{\langle 1 \rangle} }+\frac{n_{12} \big(e^{\tau \lambda _2^{\langle 1 \rangle} }-1\big)}{\lambda _2^{\langle 1 \rangle} }
        - \frac{n_{21}q_{21} \big(e^{\tau \lambda _1^{\langle 1 \rangle} }-e^{-\tau (q_{12}+q_{21}) }\big)}{\lambda _1^{\langle 1 \rangle} +q_{12}+q_{21}}-\frac{n_{22}q_{21} \big(e^{\tau \lambda_2^{\langle 1 \rangle} }-e^{-\tau(q_{12}+q_{21}) }\big)}{\lambda _2^{\langle 1 \rangle} +q_{12}+q_{21} },
    \end{align*}
    and where
    \begin{align*}
        \lambda_1^{\langle 1 \rangle}
        &= \frac{1}{{2 \alpha }} \bigg(-\kappa _1-\kappa
           _2-\alpha  q_{12}-\alpha 
           q_{21}+\sqrt{\big(\kappa _1+\kappa
           _2+\alpha  q_{12}+\alpha 
           q_{21}\big)^2-4 \big(\kappa _1
           \kappa _2+\alpha  \kappa _1
           q_{21}+\alpha  \kappa _2
           q_{12}\big)}\bigg), \\
        \lambda_2^{\langle 1 \rangle}
        &= \frac{1}{2 \alpha }\bigg(-\kappa
           _1-\kappa _2-\alpha  q_{12}-\alpha 
           q_{21}-\sqrt{\big(\kappa
           _1+\kappa _2+\alpha  q_{12}+\alpha 
           q_{21}\big)^2-4 \big(\kappa _1
           \kappa _2+\alpha  \kappa _1
           q_{21}+\alpha  \kappa _2
           q_{12}\big)}\bigg), \\
        m_{11} &= \frac{\big(\frac{\kappa_2}{\alpha}+\lambda_2^{\langle 1 \rangle}\big)\big(\lambda_1^{\langle 1 \rangle}+\frac{\kappa_2}{\alpha}+q_{21}\big)}{q_{21}\big(\lambda_2^{\langle 1 \rangle}-\lambda_1^{\langle 1 \rangle}\big)},
        \quad
        m_{12} = \frac{\big(\frac{\kappa_2}{\alpha}+\lambda_1^{\langle 1 \rangle}\big)\big(\lambda_2^{\langle 1 \rangle}+\frac{\kappa_2}{\alpha}+q_{21}\big)}{q_{21}\big(\lambda_1^{\langle 1 \rangle}-\lambda_2^{\langle 1 \rangle}\big)}, \quad
        m_{21}=\frac{\frac{\kappa_2}{\alpha}+\lambda_2^{\langle 1 \rangle}}{\lambda_2^{\langle 1 \rangle}-\lambda_1^{\langle 1 \rangle}}, \\
        m_{22} &= \frac{\frac{\kappa_2}{\alpha}+\lambda_1^{\langle 1 \rangle}}{\lambda_1^{\langle 1 \rangle}-\lambda_2^{\langle 1 \rangle}}, \quad
        n_{11}=\frac{ \gamma^{\langle 0 \rangle}_{\alpha,1} m_{11}
           q_{21}+\gamma^{\langle 0 \rangle}_{\alpha,2} m_{21}
           q_{12}}{q_{12}+q_{21}},
        \quad
        n_{12}=\frac{ \gamma^{\langle 0 \rangle}_{\alpha,1} m_{12}
           q_{21}+\gamma^{\langle 0 \rangle}_{\alpha,2} m_{22}
           q_{12}}{q_{12}+q_{21}}, \\
        n_{21} &= \frac{\gamma^{\langle 0 \rangle}_{\alpha,1} m_{11}-\gamma^{\langle 0 \rangle}_{\alpha,2} m_{21}}{q_{12}+q_{21}}, \quad
        n_{22}=\frac{\gamma^{\langle 0 \rangle}_{\alpha,1} m_{12}-\gamma^{\langle 0 \rangle}_{\alpha,2} m_{22}}{q_{12}+q_{21}},
        \quad
        \gamma_{\alpha,i}^{\langle0\rangle}=
        \frac{1}{\alpha}\left(\kappa_i\theta_i+\frac{1}{2}\sigma_i^2\left(\frac{1}{\alpha}-1\right)\right).
    \end{align*}
\end{theorem}
\begin{proof}
    We first consider the system of recursive ODEs in Theorem~\ref{thm11} when $n = 1$,
    \begin{align}
        \mathbf{\overline{u}}_\alpha^{\langle 1\rangle}(\tau)
        &= \mathbf{P}_\alpha^{\langle 1\rangle}
        \mathbf{u}_\alpha^{\langle 1\rangle}(\tau), \label{3212} \\    
        \mathbf{\overline{u}}_\alpha^{\langle 0\rangle}(\tau)
        &= \mathbf{P}_\alpha^{\langle 0\rangle}
        \mathbf{u}_\alpha^{\langle 0\rangle}(\tau)
        +\mathbf{D}_\alpha^{\langle 0\rangle}
        \mathbf{u}_\alpha^{\langle 1\rangle}(\tau), \label{3213}
    \end{align}
    with the initial conditions $\mathbf{u}_\alpha^{\langle 1\rangle}(0)=[1,1]^\top$ and $\mathbf{u}_\alpha^{\langle 0\rangle}(0)=[0,0]^\top$. We note the following eigenvalues and eigenvectors of the diagonalizable matrix $\mathbf{P}_\alpha^{\langle 1\rangle}$
    \begin{align*}
        &\lambda_1^{\langle 1 \rangle} = \frac{1}{2} \bigg(\mathrm{tr}\Big(\mathbf{P}_\alpha^{\langle 1\rangle}\Big) + \sqrt{\mathrm{tr}^2\Big(\mathbf{P}_\alpha^{\langle 1\rangle}\Big)-4\mathrm{det}\Big(\mathbf{P}_\alpha^{\langle 1\rangle}\Big)}~\bigg) \quad \text{corresponding to} \quad \mathbf{v}_1^{\langle 1 \rangle} =
        \left[ \lambda_1^{\langle 1 \rangle} + \frac{\kappa_2}{\alpha} + q_{21}, q_{21} \right]^\top, \\
        &\lambda_2^{\langle 1 \rangle} = \frac{1}{2} \bigg(\mathrm{tr}\Big(\mathbf{P}_\alpha^{\langle 1\rangle}\Big) - \sqrt{\mathrm{tr}^2\Big(\mathbf{P}_\alpha^{\langle 1\rangle}\Big)-4\mathrm{det}\Big(\mathbf{P}_\alpha^{\langle 1\rangle}\Big)}~\bigg) \quad \text{corresponding to} \quad \mathbf{v}_2^{\langle 1 \rangle} = \left[\lambda_2^{\langle 1 \rangle} + \frac{\kappa_2}{\alpha} + q_{21}, q_{21}\right]^\top, 
    \end{align*}  
    where
    $\mathrm{tr}\Big(\mathbf{P}_\alpha^{\langle 1\rangle}\Big) = -\frac{\kappa_1}{\alpha}-\frac{\kappa_2}{\alpha}-q_{12}-q_{21}$ and $
    \mathrm{det}\Big(\mathbf{P}_\alpha^{\langle 1\rangle}\Big) = \Big(\frac{\kappa_1}{\alpha}+q_{12}\Big)\Big(\frac{\kappa_2}{\alpha}+q_{21}\Big)-q_{12}q_{21}$. Consequently, we have~$\lambda_1^{\langle 1 \rangle}$ and $\lambda_2^{\langle 1 \rangle}$ defined by the theorem statement. Let $S_1=[\mathbf{v}^{\langle 1 \rangle}_1, \mathbf{v}^{\langle 1 \rangle}_2]$ 
    and $\Lambda_\alpha^{\langle 1\rangle} = \mathrm{diag}\big\{\lambda_1^{\langle 1 \rangle},\lambda_2^{\langle 1 \rangle}\big\}$. The solution of the homogeneous problem~\eqref{3212} can be found by applying the eigenvalues and eigenvectors above
    \begin{equation} \label{3214}
        \mathbf{u}_\alpha^{\langle 1\rangle}(\tau)
        = S_1e^{\tau\Lambda_\alpha^{\langle 1\rangle}}S_1^{-1}\mathbf{u}_\alpha^{\langle 1\rangle}(0)
        = \left[
        m_{11}e^{\tau\lambda_1^{\langle 1 \rangle}} + m_{12}e^{\tau\lambda_2^{\langle 1 \rangle}},
        m_{21}e^{\tau\lambda_1^{\langle 1 \rangle}} + m_{22}e^{\tau\lambda_2^{\langle 1 \rangle}}
        \right]^\top,
    \end{equation}
    where $m_{11},m_{12},m_{21}$, and $m_{22}$ are defined in the theorem statement. To investigate the nonhomogeneous problem~\eqref{3213}, we also note the simple eigenvalues and eigenvectors of the diagonalizable matrix~$\mathbf{P}_\alpha^{\langle 0\rangle}$, i.e., $\lambda_1^{\langle 0 \rangle}=0$ and $\lambda_2^{\langle 0 \rangle}=-q_{12}-q_{21}$, corresponding to $\mathbf{v}_1^{\langle 0 \rangle}=[1,1]^\top$ and $\mathbf{v}_2^{\langle 0 \rangle}=[q_{12},-q_{21}]^\top$, respectively. 
    Let $S_0=[\mathbf{v}^{\langle 0 \rangle}_1, \mathbf{v}^{\langle 0 \rangle}_2]$ 
    and $\Lambda_\alpha^{\langle 0\rangle}=\mathrm{diag}\big\{\lambda_1^{\langle 0 \rangle},\lambda_2^{\langle 0 \rangle}\big\}$. We obtain the solution of~\eqref{3213} by using~\eqref{326} and~\eqref{3214} as follows:
    \begin{align*}
        \mathbf{u}_\alpha^{\langle 0\rangle}(\tau)
        &= S_0e^{\tau\Lambda_\alpha^{\langle 0\rangle}}
        \int_0^\tau e^{-\xi\Lambda_\alpha^{\langle 0\rangle}}S_0^{-1}
        \mathbf{D}^{\langle 0 \rangle}_\alpha
        \mathbf{u}_{\alpha}^{\langle 1\rangle}(\xi)d\xi \\
        &=
        \begin{bmatrix}
            \frac{n_{11} \left(e^{\tau\lambda_1^{\langle 1 \rangle} }-1\right)}{\lambda_1^{\langle 1 \rangle}} + \frac{n_{12} \left(e^{\tau\lambda_2^{\langle 1 \rangle} }-1\right)}{\lambda_2^{\langle 1 \rangle}} + \frac{n_{21} q_{12} \left(e^{\tau\lambda_1^{\langle 1 \rangle} }-e^{-\tau(q_{12}+q_{21}) }\right)}{\lambda_1^{\langle 1 \rangle}+q_{12}+q_{21} } + \frac{n_{22} q_{12} \left(e^{\tau\lambda _2^{\langle 1 \rangle} }-e^{-\tau(q_{12}+q_{21})}\right)}{\lambda_2^{\langle 1 \rangle} +q_{12}+q_{21} } \\
            \frac{n_{11} \left( e^{\tau \lambda _1^{\langle 1 \rangle} }-1 \right)}{\lambda _1^{\langle 1 \rangle} }+\frac{n_{12} \left(e^{\tau \lambda _2^{\langle 1 \rangle}  }-1\right)}{\lambda _2^{\langle 1 \rangle} } - \frac{n_{21}q_{21} \left(e^{\tau \lambda _1^{\langle 1 \rangle} }-e^{-\tau (q_{12}+q_{21})  }\right)}{\lambda _1^{\langle 1 \rangle} +q_{12}+q_{21}}-\frac{n_{22}q_{21} \left( e^{\tau \lambda_2^{\langle 1 \rangle} } - e^{-\tau(q_{12}+q_{21}) } \right)}{\lambda_2^{\langle 1 \rangle} +q_{12}+q_{21} }
        \end{bmatrix},
    \end{align*}
    where $n_{11},n_{12},n_{21}$, and $n_{22}$ are defined in the theorem statement.
\end{proof}

\subsection{The two-state regime-switching NLD-CEV process when {$\beta\in(2,\infty)$}} \label{subsec2ty}
\medskip

This subsection follows a similar structure to the previous one, with proofs that align closely with those provided earlier. Due to their similarity, detailed proofs are omitted here, focusing instead on the main results and their applications. Before proceeding with the theorem, and following Section~\ref{subsec_hybrid__system}, let $A_\alpha^{\langle k\rangle}(\tau) := A_{\alpha, 1}^{\langle k\rangle}(\tau)$ and $B_\alpha^{\langle k\rangle}(\tau) := A_{\alpha, 2}^{\langle k\rangle}(\tau)$.
\begin{theorem} \label{thm321}
    Suppose that $R_t$ follows the system of SDEs~\eqref{model4} on $[0,T]$ with the initial conditions $R_{t} = R$ and $X_{t}=i,i \in \{1,2\}$. The conditional $\frac{n}{\alpha}$-moment for $n \in \mathbb{Z}^-_0$ is defined by
    \begin{equation*}
        \mathbf{E}^{\mathbb{Q}}\Big[ R_T^{\frac{n}{\alpha}} ~|~ R_t= R, X_t = i \Big] 
        = \mathds{1}_{ \{1\} }(i)~ \sum_{k=0}^{|n|} A_\alpha^{\langle k\rangle}(\tau)R^{-\frac{k}{\alpha}} + \mathds{1}_{ \{2\} }(i)~  \sum_{k=0}^{|n|} B_\alpha^{\langle k\rangle}(\tau)R^{-\frac{k}{\alpha}}, 
    \end{equation*}
    where $\tau=T-t\geq0$, $\mathds{1}$ is the indicator function, and $A_\alpha^{\langle k\rangle}$ and $B_\alpha^{\langle k\rangle}$ for $k=0,1,\ldots,|n|$ can be solved by the following system of
    recursive matrix differential equation:
    \begin{align}
        \mathbf{\overline{u}}_\alpha^{\langle |n|\rangle}(\tau)
        &= \mathbf{P}_\alpha^{\langle |n|\rangle}
        \mathbf{u}_\alpha^{\langle |n|\rangle}(\tau) \label{321}, \\    
        \mathbf{\overline{u}}_\alpha^{\langle \ell\rangle}(\tau)
        &= \mathbf{P}_\alpha^{\langle \ell\rangle}
        \mathbf{u}_\alpha^{\langle \ell\rangle}(\tau)
        +\mathbf{D}_\alpha^{\langle \ell\rangle}
        \mathbf{u}_\alpha^{\langle \ell+1\rangle}(\tau), \label{322} 
    \end{align}
    with the initial conditions $\mathbf{u}_\alpha^{\langle |n|\rangle}(0) = [1,1]^\top$ and $\mathbf{u}_\alpha^{\langle \ell\rangle}(0) =[0,0]^\top ~\mbox{for} ~\ell = 0, 1, \ldots, |n|-1$, where
    \begin{align*}
        &\mathbf{\overline{u}}_\alpha^{\langle k\rangle}(\tau) =
        \begin{bmatrix}
            \frac{d}{d\tau}A_\alpha^{\langle k\rangle}(\tau) \\ \frac{d}{d\tau}B_\alpha^{\langle k\rangle}(\tau)
        \end{bmatrix}, \quad
        \mathbf{P}_\alpha^{\langle k\rangle} =
        \begin{bmatrix}
            \kappa_1\frac{k}{\alpha}-q_{12}&q_{12}\\q_{21}&\kappa_2\frac{k}{\alpha}-q_{21}
        \end{bmatrix}, \quad
        \mathbf{u}_\alpha^{\langle k\rangle}(\tau)=
        \begin{bmatrix}
            A_\alpha^{\langle k\rangle}(\tau)\\B_\alpha^{\langle k\rangle}(\tau),
        \end{bmatrix}, \\
        &\mathbf{D}_\alpha^{\langle \ell\rangle} = \mathrm{diag}\left\{\overline{\gamma}_{\alpha,1}^{\langle\ell\rangle},{\overline\gamma}_{\alpha,2}^{\langle\ell\rangle}\right\}, \quad {\overline{\gamma}}_{\alpha,i}^{\langle\ell\rangle}=
        \left(\frac{\ell+1}{\alpha}\right)\left(-\kappa_i\theta_i+\frac{1}{2}\sigma_i^2\left(\frac{\ell+1}{\alpha}+1\right)\right).
    \end{align*}
\end{theorem}
\begin{proof}
    Let $m:=|n|\in\mathbb{Z}_0^+$ so that $n=-m$, and set $U_{\alpha,i}^{\langle -m\rangle}(\tau,R) := \mathbf{E}^{\mathbb{Q}}\!\left[R_T^{-m/\alpha}\mid R_t=R,\ X_t=i\right]$.
    Exactly as in Theorem~\ref{thm11} (using the regime-switching Feynman--Kac representation of Yao et al.~\cite{yao2006regime}),   $U_{\alpha,i}^{\langle -m\rangle}$ satisfies the coupled backward PDE associated with the generator of the model~\eqref{model4}, with the initial condition~$U_{\alpha,i}^{\langle -m\rangle}(0,R)=R^{-m/\alpha}$. We use the same expansion as in Theorem~\ref{thm11}, but on the negative power~basis,
    \begin{equation*}
        U_{\alpha,1}^{\langle -m\rangle}(\tau,R)=\sum_{k=0}^{m}A_\alpha^{\langle k\rangle}(\tau)R^{-\frac{k}{\alpha}},\qquad
        U_{\alpha,2}^{\langle -m\rangle}(\tau,R)=\sum_{k=0}^{m}B_\alpha^{\langle k\rangle}(\tau)R^{-\frac{k}{\alpha}}.
    \end{equation*}
    Matching $R^{-m/\alpha}$ at $\tau=0$ gives $A_\alpha^{\langle m\rangle}(0)=B_\alpha^{\langle m\rangle}(0)=1$ and $A_\alpha^{\langle \ell\rangle}(0)=B_\alpha^{\langle \ell\rangle}(0)=0$ for $\ell=0,\dots,m-1$. The only algebraic change relative to Theorem~\ref{thm11} is that
    \begin{equation*}
        \partial_R\big(R^{-k/\alpha}\big) = -\frac{k}{\alpha}R^{-k/\alpha-1},\qquad
        \partial_{RR}\big(R^{-k/\alpha}\big) = \frac{k}{\alpha}\bigg(\frac{k}{\alpha}+1\bigg)R^{-k/\alpha-2},
    \end{equation*}
    so the contribution of diffusion involves $\big(\frac{k}{\alpha}\big)\big(\frac{k}{\alpha}+1\big)$, and the drift term from the model~\eqref{model4} produces a coupling from level $\ell+1$ to $\ell$ with the coefficient
    \begin{equation*}
        \overline{\gamma}_{\alpha,i}^{\langle\ell\rangle}
        =\left(\frac{\ell+1}{\alpha}\right)\left(-\kappa_i\theta_i+\frac{1}{2}\sigma_i^2\left(\frac{\ell+1}{\alpha}+1\right)\right).
    \end{equation*}
    Substituting the expansion into the coupled PDE and collecting the coefficients of the linearly independent powers $\{R^{-k/\alpha}\}_{k=0}^{m}$ yields the ODE system \eqref{321} and \eqref{322} in vector form with $\mathbf{u}_\alpha^{\langle k\rangle}=[A_\alpha^{\langle k\rangle},\,B_\alpha^{\langle k\rangle}]^\top$,
    $\mathbf{P}_\alpha^{\langle k\rangle}$, and $\mathbf{D}_\alpha^{\langle \ell\rangle}$ exactly as stated.
\end{proof}

The next theorem addresses the scenario where the parameters are treated as constant functions,~simplifying the model\rq s structure and allowing for direct analytical solutions under these fixed~conditions.
\begin{theorem}\label{thm7}
    Suppose that $R_t$ follows the system of SDEs~\eqref{model4} such that $\kappa_1 = \kappa_2 = \kappa, \theta_1 = \theta_2 = \theta$ and~$\sigma_1 = \sigma_2 = \sigma$ on $[0,T]$ with the initial conditions $R_{t}=R$ and $X_{t}=i,i\in\{1,2\}$. The conditional~$\frac{n}{\alpha}$-moment for $n\in\mathbb{Z}^-_0$ is defined by
    \begin{align*}
        \mathbf{E}^{\mathbb{Q}}\Big[ R_T^{\frac{n}{\alpha}} ~|~ R_t= R, X_t = i \Big] = \sum_{k=0}^{|n|} \left(\frac{e^{\frac{|n|\kappa\tau}{\alpha }}}{(|n|-k)!}\left(\frac{\alpha-\alpha e^{-\frac{\kappa\tau}{\alpha}}}{\kappa}\right)^{|n|-k}
        \left(\prod_{j=1}^{|n|-k} \overline{\gamma}_\alpha^{\langle |n|-j \rangle}
        \right)\right)R^{-\frac{k}{\alpha}},
    \end{align*}
    where the product\rq s term equals $1$ when $k=|n|$.
\end{theorem}
\begin{proof}
    The proof follows the same reduction as in Theorem~\ref{thm312}, now starting from the negative-moment~ODE chain~\eqref{321} and \eqref{322} (i.e., Theorem~\ref{thm321}) under the parameter restrictions $\kappa_1=\kappa_2=\kappa$, $\theta_1=\theta_2=\theta$, and $\sigma_1=\sigma_2=\sigma$.
    In this case
    \begin{equation*}
        \mathbf{P}_\alpha^{\langle k\rangle}=
        \begin{bmatrix}
            \kappa\frac{k}{\alpha}-q_{12} & q_{12}\\
            q_{21} & \kappa\frac{k}{\alpha}-q_{21}
        \end{bmatrix},
        \qquad
        \mathbf{D}_\alpha^{\langle \ell\rangle}
        =\overline{\gamma}_\alpha^{\langle \ell\rangle}\,\mathbf{I}_2,
    \end{equation*}
    where $\overline{\gamma}_\alpha^{\langle \ell\rangle}$ is the common value of $\overline{\gamma}_{\alpha,1}^{\langle \ell\rangle}=\overline{\gamma}_{\alpha,2}^{\langle \ell\rangle}$. Again $\mathrm{span}\{[1,1]^\top\}$ is invariant because $\mathbf{P}_\alpha^{\langle k\rangle}[1,1]^\top=\kappa\frac{k}{\alpha}[1,1]^\top$, and the forcing preserves this subspace since $\mathbf{D}_\alpha^{\langle \ell\rangle}$ is a scalar multiple of $\mathbf{I}_2$. Hence, $\mathbf{u}_\alpha^{\langle k\rangle}(\tau)=d_k(\tau)[1,1]^\top$, and $A_\alpha^{\langle k\rangle}(\tau)=B_\alpha^{\langle k\rangle}(\tau)=d_k(\tau)$, so the moment is independent of $i$.
    
    Let $m:=|n|$. The scalar chain becomes
    \begin{equation*}
        d_m'(\tau)=\kappa\frac{m}{\alpha}d_m(\tau),\qquad d_m(0)=1,
    \end{equation*}
    and for $\ell=0,1,\dots,m-1$, we have
    \begin{equation*}
        d_\ell'(\tau)=\kappa\frac{\ell}{\alpha}d_\ell(\tau)+\overline{\gamma}_\alpha^{\langle \ell\rangle}d_{\ell+1}(\tau),
        \qquad d_\ell(0)=0.
    \end{equation*}
    Define the following for $k = 0, 1, \dots, m$:
    \begin{equation*}
        d_k(\tau):=\frac{e^{\frac{m\kappa\tau}{\alpha}}}{(m-k)!}
        \left(\frac{\alpha-\alpha e^{-\frac{\kappa\tau}{\alpha}}}{\kappa}\right)^{m-k}
        \left(\prod_{j=1}^{m-k}\overline{\gamma}_\alpha^{\langle m-j\rangle}\right),
    \end{equation*}
    with the empty product equal to $1$ when $k=m$. Then $d_m(\tau)=e^{\frac{m\kappa\tau}{\alpha}}$ solves the top equation. For~$k<m$, one checks by differentiation that $d_k'(\tau)=\kappa\frac{k}{\alpha}d_k(\tau)+\overline{\gamma}_\alpha^{\langle k\rangle}d_{k+1}(\tau)$, while $d_k(0)=0$ for $k<m$ because $\left(1-e^{-\frac{\kappa\tau}{\alpha}}\right)^{m-k}$ vanishes at $\tau=0$. Substituting $A_\alpha^{\langle k\rangle}=B_\alpha^{\langle k\rangle}=d_k$ into the expansion in Theorem~\ref{thm321} yields the stated closed form.
\end{proof}

The following Theorem is to illustrate the use of Theorem~\ref{thm321} in the case of~the conditional~$-\frac{1}{\alpha}$-moment.
\begin{theorem} \label{thm323}
    Suppose that $R_t$ follows the system of SDEs~\eqref{model4} on $[0,T]$ with the initial conditions $R_{t}=R$ and $X_{t}=i,i\in\{1,2\}$. The conditional $-\frac{1}{\alpha}$-moment is defined by
    \begin{align}
        \mathbf{E}^{\mathbb{Q}}\Big[ R_T^{-\frac{1}{\alpha}} ~|~ R_t= R, X_t = i \Big] 
        = \mathds{1}_{ \{1\} }(i)
        \left(A_\alpha^{\langle 0\rangle}(\tau)+A_\alpha^{\langle 1\rangle}(\tau)R^{-\frac{1}{\alpha}}\right)
        +\mathds{1}_{ \{2\} }(i)
        \left(B_\alpha^{\langle 0\rangle}(\tau)+B_\alpha^{\langle 1\rangle}(\tau)R^{-\frac{1}{\alpha}}\right),
    \end{align}
    where $\tau = T-t \geq 0$, $\mathds{1}$ is the indicator function, and $A_\alpha^{\langle k\rangle}$ and $B_\alpha^{\langle k\rangle}$ for $k=0,1$ can be calculated by the~following:
    \begin{align*}
        A_\alpha^{\langle 1\rangle}(\tau)
        &= m_{11}e^{\tau\lambda_1^{\langle 1 \rangle}} + m_{12}e^{\tau\lambda_2^{\langle 1 \rangle}}, \quad B_\alpha^{\langle 1\rangle}(\tau)
        = m_{21}e^{\tau\lambda_1^{\langle 1 \rangle}} + m_{22}e^{\tau\lambda_2^{\langle 1 \rangle}}, \\
        A_\alpha^{\langle 0\rangle}(\tau)
        &= \frac{n_{11} \big(e^{\tau\lambda_1^{\langle 1 \rangle} 
           }-1\big)}{\lambda_1^{\langle 1 \rangle}} + \frac{n_{12}
           \big(e^{\tau\lambda_2^{\langle 1 \rangle}  }-1 \big)}{\lambda
           _2^{\langle 1 \rangle}}
           + \frac{n_{21} q_{12} \big(e^{\tau\lambda _1^{\langle 1 \rangle}}-e^{-\tau(q_{12}+q_{21}) 
           }\big)}{\lambda
           _1^{\langle 1 \rangle}+q_{12}+q_{21} }
           + \frac{n_{22} q_{12} \big(e^{\tau\lambda _2^{\langle 1 \rangle}}-e^{-\tau(q_{12}+q_{21})
           }\big)}{\lambda_2^{\langle 1 \rangle} +q_{12}+q_{21} }, \\
        B_\alpha^{\langle 0\rangle}(\tau)
        &= \frac{n_{11} \big(e^{\tau \lambda _1^{\langle 1 \rangle}  }-1\big)}{\lambda _1^{\langle 1 \rangle} }+\frac{n_{12} \big(e^{\tau \lambda _2^{\langle 1 \rangle}  }-1\big)}{\lambda _2^{\langle 1 \rangle} }
        - \frac{n_{21}q_{21} \big(e^{\tau \lambda _1^{\langle 1 \rangle}  }-e^{-\tau (q_{12}+q_{21})  }\big)}{\lambda _1^{\langle 1 \rangle} +q_{12}+q_{21}}-\frac{n_{22}q_{21} \big(e^{\tau \lambda_2^{\langle 1 \rangle}  }-e^{-\tau(q_{12}+q_{21})  }\big)}{\lambda _2^{\langle 1 \rangle} +q_{12}+q_{21} },
    \end{align*}
    and where
    \begin{align*}
        \lambda_1^{\langle 1 \rangle}
        &= \frac{1}{{2 \alpha }} \bigg(\kappa _1+\kappa
           _2-\alpha  q_{12}-\alpha 
           q_{21}+\sqrt{\big(-\kappa _1-\kappa
           _2+\alpha  q_{12}+\alpha 
           q_{21}\big)^2-4 \big(\kappa _1
           \kappa _2-\alpha  \kappa _1
           q_{21}-\alpha  \kappa _2
           q_{12}\big)} \bigg), \\
        \lambda_2^{\langle 1 \rangle}
        &= \frac{1}{{2 \alpha }}\bigg(\kappa _1+\kappa
           _2-\alpha  q_{12}-\alpha 
           q_{21}-\sqrt{\big(-\kappa _1-\kappa
           _2+\alpha  q_{12}+\alpha 
           q_{21}\big)^2-4 \big(\kappa _1
           \kappa _2-\alpha  \kappa _1
           q_{21}-\alpha  \kappa _2
           q_{12}\big)}\bigg), \\
        m_{11} &= \frac{\big(-\frac{\kappa_2}{\alpha}+\lambda_2^{\langle 1 \rangle}\big)\big(\lambda_1^{\langle 1 \rangle}-\frac{\kappa_2}{\alpha}+q_{21}\big)}{q_{21}\big(\lambda_2^{\langle 1 \rangle}-\lambda_1^{\langle 1 \rangle}\big)},
        \quad
        m_{12} = \frac{\big(-\frac{\kappa_2}{\alpha}+\lambda_1^{\langle 1 \rangle}\big)\big(\lambda_2^{\langle 1 \rangle}-\frac{\kappa_2}{\alpha}+q_{21}\big)}{q_{21}\big(\lambda_1^{\langle 1 \rangle}-\lambda_2^{\langle 1 \rangle}\big)},
        \quad
        m_{21}=\frac{\frac{\kappa_2}{\alpha}-\lambda_2^{\langle 1 \rangle}}{\lambda_1^{\langle 1 \rangle}-\lambda_2^{\langle 1 \rangle}}, \\
        m_{22} &= \frac{\frac{\kappa_2}{\alpha}-\lambda_1^{\langle 1 \rangle}}{\lambda_2^{\langle 1 \rangle}-\lambda_1^{\langle 1 \rangle}}, \quad
        n_{11}=\frac{ \overline{\gamma}^{\langle 0 \rangle}_{\alpha,1} m_{11}
           q_{21}+\overline{\gamma}^{\langle 0 \rangle}_{\alpha,2} m_{21}
           q_{12}}{q_{12}+q_{21}},
        \quad
        n_{12}=\frac{ \overline{\gamma}^{\langle 0 \rangle}_{\alpha,1} m_{12}
           q_{21}+\overline{\gamma}^{\langle 0 \rangle}_{\alpha,2} m_{22}
           q_{12}}{q_{12}+q_{21}}, \\
        n_{21} &= \frac{\overline{\gamma}^{\langle 0 \rangle}_{\alpha,1} m_{11}-\overline{\gamma}^{\langle 0 \rangle}_{\alpha,2} m_{21}}{q_{12}+q_{21}}, \quad
        n_{22}=\frac{\overline{\gamma}^{\langle 0 \rangle}_{\alpha,1} m_{12}-\overline{\gamma}^{\langle 0 \rangle}_{\alpha,2} m_{22}}{q_{12}+q_{21}},
        \quad
        \overline{\gamma}_{\alpha,i}^{\langle0\rangle}=
        \frac{1}{\alpha}\left(-\kappa_i\theta_i+\frac{1}{2}\sigma_i^2\left(\frac{1}{\alpha}+1\right)\right).
    \end{align*}
\end{theorem}
\begin{proof}
    This is the $|n|=1$ specialization of the negative-moment recursion in Theorem~\ref{thm321}. Let $\mathbf{u}_\alpha^{\langle 1\rangle}(\tau)=[A_\alpha^{\langle 1\rangle}(\tau),\,B_\alpha^{\langle 1\rangle}(\tau)]^\top$ and $\mathbf{u}_\alpha^{\langle 0\rangle}(\tau)=[A_\alpha^{\langle 0\rangle}(\tau),\,B_\alpha^{\langle 0\rangle}(\tau)]^\top$. They satisfy
    \begin{equation*}
        \frac{d}{d\tau}\mathbf{u}_\alpha^{\langle 1\rangle}(\tau)
        =\mathbf{P}_\alpha^{\langle 1\rangle}\mathbf{u}_\alpha^{\langle 1\rangle}(\tau),
        \qquad
        \frac{d}{d\tau}\mathbf{u}_\alpha^{\langle 0\rangle}(\tau)
        =\mathbf{P}_\alpha^{\langle 0\rangle}\mathbf{u}_\alpha^{\langle 0\rangle}(\tau)
        +\mathbf{D}_\alpha^{\langle 0\rangle}\mathbf{u}_\alpha^{\langle 1\rangle}(\tau),
    \end{equation*}
    with the initial conditions $\mathbf{u}_\alpha^{\langle 1\rangle}(0)=[1,1]^\top$ and $\mathbf{u}_\alpha^{\langle 0\rangle}(0)=[0,0]^\top$,
    where now $\mathbf{P}_\alpha^{\langle k\rangle}$ is the matrix in~Theorem~\ref{thm323}~(coming from Model~\eqref{model4}) and $\mathbf{D}_\alpha^{\langle 0\rangle}=\mathrm{diag}\{\overline{\gamma}_{\alpha,1}^{\langle0\rangle},\overline{\gamma}_{\alpha,2}^{\langle0\rangle}\}$.
    
    The constant matrix $\mathbf{P}_\alpha^{\langle 1\rangle}$ is diagonalizable with eigenvalues $\lambda_1^{\langle 1\rangle},\lambda_2^{\langle 1\rangle}$ given in the theorem statement. Hence,
    \begin{equation*}
        \mathbf{u}_\alpha^{\langle 1\rangle}(\tau)
        =e^{\tau\mathbf{P}_\alpha^{\langle 1\rangle}}\mathbf{u}_\alpha^{\langle 1\rangle}(0)
        =
        \begin{bmatrix}
        m_{11}e^{\tau\lambda_1^{\langle 1\rangle}}+m_{12}e^{\tau\lambda_2^{\langle 1\rangle}}\\[2pt]
        m_{21}e^{\tau\lambda_1^{\langle 1\rangle}}+m_{22}e^{\tau\lambda_2^{\langle 1\rangle}}
        \end{bmatrix},
    \end{equation*}
    with $m_{ij}$ as stated. For $\mathbf{u}_\alpha^{\langle 0\rangle}$, the variation-of-constants formula yields
    \begin{equation*}
        \mathbf{u}_\alpha^{\langle 0\rangle}(\tau)
        =\int_0^\tau e^{(\tau-\xi)\mathbf{P}_\alpha^{\langle 0\rangle}}\,
        \mathbf{D}_\alpha^{\langle 0\rangle}\mathbf{u}_\alpha^{\langle 1\rangle}(\xi)\,d\xi.
    \end{equation*}
    Here, $\mathbf{P}_\alpha^{\langle 0\rangle}=\bigl[\begin{smallmatrix}-q_{12}&q_{12}\\ q_{21}&-q_{21}\end{smallmatrix}\bigr]$ is the same as in Theorem~\ref{thm313}, so it has the eigenpairs $(0,[1,1]^\top)$ and $(-(q_{12}+q_{21}),[q_{12},-q_{21}]^\top)$. Decomposing the forcing term
    $\mathbf{D}_\alpha^{\langle 0\rangle}\mathbf{u}_\alpha^{\langle 1\rangle}(\xi)
    =\bigl[\overline{\gamma}_{\alpha,1}^{\langle0\rangle}A_\alpha^{\langle 1\rangle}(\xi),\,
    \overline{\gamma}_{\alpha,2}^{\langle0\rangle}B_\alpha^{\langle 1\rangle}(\xi)\bigr]^\top$ on this basis gives the scalar coefficients
    \begin{equation*}
a(\xi)=\frac{\overline{\gamma}_{\alpha,1}^{\langle0\rangle}A_\alpha^{\langle 1\rangle}(\xi)\,q_{21}
        +\overline{\gamma}_{\alpha,2}^{\langle0\rangle}B_\alpha^{\langle 1\rangle}(\xi)\,q_{12}}{q_{12}+q_{21}},
        \qquad
        b(\xi)=\frac{\overline{\gamma}_{\alpha,1}^{\langle0\rangle}A_\alpha^{\langle 1\rangle}(\xi)
        -\overline{\gamma}_{\alpha,2}^{\langle0\rangle}B_\alpha^{\langle 1\rangle}(\xi)}{q_{12}+q_{21}},
    \end{equation*}
    so that $a(\xi)=n_{11}e^{\xi\lambda_1^{\langle 1\rangle}}+n_{12}e^{\xi\lambda_2^{\langle 1\rangle}}$ and $b(\xi)=n_{21}e^{\xi\lambda_1^{\langle 1\rangle}}+n_{22}e^{\xi\lambda_2^{\langle 1\rangle}}$, with $n_{ij}$ as stated. Evaluating the resulting elementary integrals
    \begin{equation*}
        \int_0^\tau e^{\xi\lambda}\,d\xi=\frac{e^{\tau\lambda}-1}{\lambda},\qquad
        \int_0^\tau e^{-(q_{12}+q_{21})(\tau-\xi)}e^{\xi\lambda}\,d\xi
        =\frac{e^{\tau\lambda}-e^{-\tau(q_{12}+q_{21})}}{\lambda+q_{12}+q_{21}}
    \end{equation*}
    yields the closed forms for $A_\alpha^{\langle 0\rangle}(\tau)$ and $B_\alpha^{\langle 0\rangle}(\tau)$ in the theorem statement. Substituting these coefficients into the $|n|=1$ expansion in Theorem~\ref{thm321} gives~\eqref{thm323}.
\end{proof}


\subsection{Unconditional moments}
\medskip

On the basis of the conditions established in Sections~\ref{subsec02} and~\ref{subsec2ty}, this section introduces two principal theorems for the unconditional moments. The formulas for unconditional moments corresponding to results of Theorems~\ref{thm313} and~\ref{thm323} are obtained by taking $\tau \rightarrow \infty$. Notably, the resulting formulas no longer depend on the initial value $R$ or the initial state $X_t$, which are simplified into two-term summations as demonstrated below.
\begin{theorem} \label{thm_uncon_1}
    Keeping the notation in Theorem~\ref{thm313}, we have
    \begin{equation*}
        \lim_{\tau\rightarrow\infty} U_{\alpha,i}^{\langle 1\rangle}(\tau,R) = \lim_{T\rightarrow\infty} \mathbf{E}^{\mathbb{Q}} \Big[ R_T^{\frac{1}{\alpha}} ~|~ R_t= R, X_t = i \Big]         = \frac{-n_{11}}{\lambda_1^{\langle 1 \rangle}}+\frac{-n_{12}}{\lambda_2^{\langle 1 \rangle}},
    \end{equation*}
    where $i$ is in the state space $\mathcal{M}_2 =\{1,2\}$, and
    \begin{align*}
        \lambda_1^{\langle 1 \rangle}
        &= \frac{1}{{2 \alpha }} \bigg(-\kappa _1-\kappa
           _2-\alpha  q_{12}-\alpha 
           q_{21}+\sqrt{\big(\kappa _1+\kappa
           _2+\alpha  q_{12}+\alpha 
           q_{21}\big)^2-4 \big(\kappa _1
           \kappa _2+\alpha  \kappa _1
           q_{21}+\alpha  \kappa _2
           q_{12}\big)}\bigg), \\
        \lambda_2^{\langle 1 \rangle}
        &= \frac{1}{2 \alpha } \bigg(-\kappa
           _1-\kappa _2-\alpha  q_{12}-\alpha 
           q_{21}-\sqrt{\big(\kappa
           _1+\kappa _2+\alpha  q_{12}+\alpha 
           q_{21}\big)^2-4 \big(\kappa _1
           \kappa _2+\alpha  \kappa _1
           q_{21}+\alpha  \kappa _2
           q_{12}\big)}\bigg), \\
        n_{11} &=\frac{ \gamma^{\langle 0 \rangle}_{\alpha,1} m_{11}
           q_{21}+\gamma^{\langle 0 \rangle}_{\alpha,2} m_{21}
           q_{12}}{q_{12}+q_{21}},
        \quad
        n_{12}=\frac{ \gamma^{\langle 0 \rangle}_{\alpha,1} m_{12}
           q_{21}+\gamma^{\langle 0 \rangle}_{\alpha,2} m_{22}
           q_{12}}{q_{12}+q_{21}}, \\
        m_{11} &= \frac{\big(\frac{\kappa_2}{\alpha}+\lambda_2^{\langle 1 \rangle}\big)\big(\lambda_1^{\langle 1 \rangle}+\frac{\kappa_2}{\alpha}+q_{21}\big)}{q_{21}\big(\lambda_2^{\langle 1 \rangle}-\lambda_1^{\langle 1 \rangle}\big)},
        \quad
        m_{12} = \frac{\big(\frac{\kappa_2}{\alpha}+\lambda_1^{\langle 1 \rangle}\big)\big(\lambda_2^{\langle 1 \rangle}+\frac{\kappa_2}{\alpha}+q_{21}\big)}{q_{21}\big(\lambda_1^{\langle 1 \rangle}-\lambda_2^{\langle 1 \rangle}\big)}, \\
        m_{21} &=\frac{\frac{\kappa_2}{\alpha}+\lambda_2^{\langle 1 \rangle}}{\lambda_2^{\langle 1 \rangle}-\lambda_1^{\langle 1 \rangle}}, \quad m_{22} = \frac{\frac{\kappa_2}{\alpha}+\lambda_1^{\langle 1 \rangle}}{\lambda_1^{\langle 1 \rangle}-\lambda_2^{\langle 1 \rangle}}, \quad \gamma_{\alpha,i}^{\langle0\rangle} =
        \frac{1}{\alpha}\left(\kappa_i\theta_i+\frac{1}{2}\sigma_i^2\left(\frac{1}{\alpha}-1\right)\right).
    \end{align*}
\end{theorem}
\begin{proof}
    We have to only verify that the eigenvalues $\lambda_1^{\langle 1 \rangle}$ and $\lambda_2^{\langle 1 \rangle}$ are negative. To demonstrate that $\lambda_1^{\langle 1 \rangle}$ is always negative under the given conditions, we simplify the expressions by defining
    \begin{equation*}
        A := \kappa_1 + \kappa_2 + \alpha q_{12} + \alpha q_{21} \quad \text{and} \quad B := \kappa_1 \kappa_2 + \alpha \kappa_1 q_{21} + \alpha \kappa_2 q_{12}.
    \end{equation*}
    Given that all parameters $\kappa_1$, $\kappa_2$, $\alpha$, $q_{12}$, and $q_{21}$ are positive real numbers, it follows that $A > 0$ and~$B > 0$. The discriminant in the expression for $\lambda_1^{\langle 1 \rangle}$ becomes $\Delta := A^2 - 4B$. Substituting these definitions back into the expression for $\lambda_1^{\langle 1 \rangle}$, we have $\lambda_1^{\langle 1 \rangle} = \frac{1}{2\alpha} \big( -A + \sqrt{\Delta} \big)$.

    Since $B > 0$, it implies that $\Delta = A^2 - 4B < A^2$, which leads to $\sqrt{\Delta} < A$. This inequality indicates that the numerator $-A + \sqrt{\Delta}$ is negative because $-A + \sqrt{\Delta} < -A + A = 0$. Therefore, $\lambda_1^{\langle 1 \rangle}$ is negative. Similarly, the case of $\lambda_2^{\langle 1 \rangle}$ is not difficult to verify, and the proof is omitted. Since the eigenvalues $\lambda_1^{\langle 1 \rangle}$ and $\lambda_2^{\langle 1 \rangle}$ are negative, taking $\tau \to \infty$ yields $A_\alpha^{\langle 1\rangle}(\tau) \to 0$, $B_\alpha^{\langle 1\rangle}(\tau) \to 0$, and we have
    \begin{equation*}
        \lim_{T\rightarrow\infty} \mathbf{E}^{\mathbb{Q}} \Big[ R_T^{\frac{1}{\alpha}} ~|~ R_t= R, X_t = i \Big] = \mathds{1}_{ \{1\} }(i) A_\alpha^{\langle 0\rangle}(\tau)
        +\mathds{1}_{ \{2\} }(i) B_\alpha^{\langle 0\rangle}(\tau),
    \end{equation*}
    where $A_\alpha^{\langle 0\rangle}(\tau) \to \frac{-n_{11}}{\lambda_1^{\langle 1 \rangle}}+\frac{-n_{12}}{\lambda_2^{\langle 1 \rangle}}$ \ and \ $B_\alpha^{\langle 0\rangle}(\tau) \to \frac{-n_{11}}{\lambda_1^{\langle 1 \rangle}}+\frac{-n_{12}}{\lambda_2^{\langle 1 \rangle}}$, which converges to the same value. Since both expressions are equal, the indicator function, which differentiates between the two cases, is not needed in this context. Therefore, this completes the proof as required.
\end{proof}

In the context of the NLD-CEV process~\eqref{main_process<2}, by applying the given assumptions that $\kappa_1=\kappa_2=\kappa, \theta_1=\theta_2=\theta$, and $\sigma_1=\sigma_2=\sigma$, we have simplified the complex expression to a form that is independent of the transition rates $q_{12}$ and $q_{21}$. The final result depends only on the parameters $\kappa$, $\theta$, $\sigma$, and $\alpha$, providing a more tractable expression for further analysis.
\begin{corollary} \label{cor_uncon_1}
    According to Theorem~\ref{thm_uncon_1} and the system~\eqref{model3} with $\kappa_1=\kappa_2=\kappa, \theta_1=\theta_2=\theta$, and~$\sigma_1=\sigma_2=\sigma$, we have
    \begin{equation*}
        \lim_{\tau\rightarrow\infty} U_{\alpha,i}^{\langle 1\rangle}(\tau,R) = \theta + \frac{\sigma^2}{2 \kappa} \left( \frac{1}{\alpha} - 1 \right),
    \end{equation*}
\end{corollary}
\begin{proof}
    Under the assumptions that $\kappa_1 = \kappa_2 = \kappa$, $\theta_1 = \theta_2 = \theta$, and $\sigma_1 = \sigma_2 = \sigma$, our aim is to simplify the expression presented in Theorem~\ref{thm_uncon_1} to eliminate the terms $q_{12}$ and $q_{21}$, expressing the result solely in terms of $\kappa$, $\theta$, $\sigma$, and $\alpha$. First, we simplify the eigenvalues $\lambda_1^{\langle 1 \rangle}$ and $\lambda_2^{\langle 1 \rangle}$. Given the parameters, the eigenvalues become
    \begin{equation*}
        \lambda_1^{\langle 1 \rangle} = \frac{1}{2\alpha} \left( -2\kappa - \alpha(q_{12} + q_{21}) + \sqrt{\Delta} \right) \quad \text{and} \quad \lambda_2^{\langle 1 \rangle} = \frac{1}{2\alpha} \left( -2\kappa - \alpha(q_{12} + q_{21}) - \sqrt{\Delta} \right),
    \end{equation*}
    where $\Delta := \left( 2\kappa + \alpha(q_{12} + q_{21}) \right)^2 - 4\left( \kappa^2 + \alpha\kappa(q_{12} + q_{21}) \right)$ denotes the discriminant. Simplifying the discriminant $\Delta$, we find
    \begin{equation*}
        \sqrt{\Delta} = \alpha(q_{12} + q_{21}).
    \end{equation*}  
    Substituting back into the expressions for the eigenvalues, we have
    \begin{equation*}
        \lambda_1^{\langle 1 \rangle} = \frac{-\kappa}{\alpha}, \quad \lambda_2^{\langle 1 \rangle} = \frac{-\kappa}{\alpha} - (q_{12} + q_{21}), \quad \text{and} \quad \lambda_1^{\langle 1 \rangle} - \lambda_2^{\langle 1 \rangle} = q_{12} + q_{21}.
    \end{equation*}
    We proceed to simplify the coefficients $m_{ij}$, and we have
    \begin{equation*}
        \frac{\kappa}{\alpha} + \lambda_2^{\langle 1 \rangle} = - (q_{12} + q_{21}) \quad \text{and} \quad \lambda_1^{\langle 1 \rangle} + \frac{\kappa}{\alpha} + q_{21} = q_{21}.
    \end{equation*}
    Therefore, the coefficient $m_{11}$ simplifies to
    \begin{equation*}
        m_{11} = \frac{\big(\frac{\kappa_2}{\alpha}+\lambda_2^{\langle 1 \rangle}\big)\big(\lambda_1^{\langle 1 \rangle}+\frac{\kappa_2}{\alpha}+q_{21}\big)}{q_{21}\big(\lambda_2^{\langle 1 \rangle}-\lambda_1^{\langle 1 \rangle}\big)} = \frac{ - (q_{12} + q_{21}) q_{21} }{ -q_{21} (q_{12} + q_{21}) } = 1.
    \end{equation*}
    Similarly, we find that $m_{12} = 0$, $m_{21} = 1$, and $m_{22} = 0$. We then simplify $n_{11}$ and $n_{12}$. Under the assumptions, we have $\gamma_{\alpha,1}^{\langle 0 \rangle} = \gamma_{\alpha,2}^{\langle 0 \rangle} =: \gamma_{\alpha}^{\langle 0 \rangle}$ and
    \begin{equation*}
        n_{11} = \frac{ \gamma_{\alpha}^{\langle 0 \rangle} ( m_{11} q_{21} + m_{21} q_{12} ) }{ q_{12} + q_{21} } = \gamma_{\alpha}^{\langle 0 \rangle} \quad \text{and} \quad n_{12} = 0.
    \end{equation*}
    Substituting the simplified values into the original expression, we obtain
    \begin{equation*}
        \frac{ - n_{11} }{ \lambda_1^{\langle 1 \rangle} } + \frac{ - n_{12} }{ \lambda_2^{\langle 1 \rangle} } = \frac{ - \gamma_{\alpha}^{\langle 0 \rangle} }{ - \kappa / \alpha } + 0 = \gamma_{\alpha}^{\langle 0 \rangle} \cdot \frac{ \alpha }{ \kappa } = \frac{1}{\alpha} \left( \kappa \theta + \frac{1}{2} \sigma^2 \left( \frac{1}{\alpha} - 1 \right) \right) \cdot \frac{ \alpha }{ \kappa } = \theta + \frac{ \sigma^2 }{ 2\kappa } \left( \frac{1}{\alpha} - 1 \right). \qedhere
    \end{equation*}
\end{proof}
Building upon the previous part, this part is developed in a similar manner. The proofs are analogous to those provided earlier and will be omitted where appropriate. They are straightforward counterparts to Theorem~\ref{thm_uncon_1} and Corollary~\ref{cor_uncon_1}, and the formulas are presented in the following results.
\begin{theorem} \label{thm_uncon_2}
    Keeping the notation in Theorem~\ref{thm323}, we have
    \begin{align}
        \lim_{\tau\rightarrow\infty} U_{\alpha,i}^{\langle -1\rangle}(\tau,R) = \mathbf{E}^{\mathbb{Q}}\Big[ R_T^{-\frac{1}{\alpha}} ~|~ R_t= R, X_t = i \Big] 
        = \frac{-n_{11}}{\lambda_1^{\langle 1 \rangle} } + \frac{-n_{12}}{\lambda_2^{\langle 1 \rangle} },
    \end{align}
    where $i$ is in the state space $\mathcal{M}_2 =\{1,2\}$, and
    \begin{align*}
        \lambda_1^{\langle 1 \rangle}
        &= \frac{1}{{2 \alpha }} \bigg(\kappa _1+\kappa
           _2-\alpha  q_{12}-\alpha 
           q_{21}+\sqrt{\big(-\kappa _1-\kappa
           _2+\alpha  q_{12}+\alpha 
           q_{21}\big)^2-4 \big(\kappa _1
           \kappa _2-\alpha  \kappa _1
           q_{21}-\alpha  \kappa _2
           q_{12}\big)}\bigg), \\
        \lambda_2^{\langle 1 \rangle}
        &= \frac{1}{{2 \alpha }} \bigg(\kappa _1+\kappa
           _2-\alpha  q_{12}-\alpha 
           q_{21}-\sqrt{\big(-\kappa _1-\kappa
           _2+\alpha  q_{12}+\alpha 
           q_{21}\big)^2-4 \big(\kappa _1
           \kappa _2-\alpha  \kappa _1
           q_{21}-\alpha  \kappa _2
           q_{12}\big)}\bigg), \\
        n_{11} &= \frac{ \overline{\gamma}^{\langle 0 \rangle}_{\alpha,1} m_{11}
           q_{21}+\overline{\gamma}^{\langle 0 \rangle}_{\alpha,2} m_{21}
           q_{12}}{q_{12}+q_{21}},
        \quad
        n_{12}=\frac{ \overline{\gamma}^{\langle 0 \rangle}_{\alpha,1} m_{12}
           q_{21}+\overline{\gamma}^{\langle 0 \rangle}_{\alpha,2} m_{22}
           q_{12}}{q_{12}+q_{21}}, \\
        m_{11} &= \frac{\big(-\frac{\kappa_2}{\alpha}+\lambda_2^{\langle 1 \rangle}\big)\big(\lambda_1^{\langle 1 \rangle}-\frac{\kappa_2}{\alpha}+q_{21}\big)}{q_{21}\big(\lambda_2^{\langle 1 \rangle}-\lambda_1^{\langle 1 \rangle}\big)},
        \quad
        m_{12}=\frac{\big(-\frac{\kappa_2}{\alpha}+\lambda_1^{\langle 1 \rangle}\big)\big(\lambda_2^{\langle 1 \rangle}-\frac{\kappa_2}{\alpha}+q_{21}\big)}{q_{21}\big(\lambda_1^{\langle 1 \rangle}-\lambda_2^{\langle 1 \rangle}\big)}, \\
        m_{21} &= \frac{\frac{\kappa_2}{\alpha}-\lambda_2^{\langle 1 \rangle}}{\lambda_1^{\langle 1 \rangle}-\lambda_2^{\langle 1 \rangle}}, \quad
        m_{22} = \frac{\frac{\kappa_2}{\alpha}-\lambda_1^{\langle 1 \rangle}}{\lambda_2^{\langle 1 \rangle}-\lambda_1^{\langle 1 \rangle}}, \quad \overline{\gamma}_{\alpha,i}^{\langle0\rangle} =
        \frac{1}{\alpha}\left(-\kappa_i\theta_i+\frac{1}{2}\sigma_i^2\left(\frac{1}{\alpha}+1\right)\right).
    \end{align*}
\end{theorem}
\begin{proof}
    Since $\tau=T-t$, the limits $T\to\infty$ and $\tau\to\infty$ are equivalent. From Theorem~\ref{thm323}, we have, for each $i\in\{1,2\}$, the representation
    \begin{equation*}
        U_{\alpha,i}^{\langle -1\rangle}(\tau,R)
        =\mathds{1}_{\{1\}}(i)\Big(A_\alpha^{\langle 0\rangle}(\tau)+A_\alpha^{\langle 1\rangle}(\tau)R^{-1/\alpha}\Big)
        +\mathds{1}_{\{2\}}(i)\Big(B_\alpha^{\langle 0\rangle}(\tau)+B_\alpha^{\langle 1\rangle}(\tau)R^{-1/\alpha}\Big),
    \end{equation*}
    where $A_\alpha^{\langle 1\rangle}(\tau)$ and $B_\alpha^{\langle 1\rangle}(\tau)$ are linear combinations of $e^{\tau\lambda_1^{\langle 1\rangle}}$ and $e^{\tau\lambda_2^{\langle 1\rangle}}$, and $A_\alpha^{\langle 0\rangle}(\tau)$ and $B_\alpha^{\langle 0\rangle}(\tau)$ are the sums of terms of the forms
    \begin{equation*}
        \frac{e^{\tau\lambda}-1}{\lambda}\qquad\text{and}\qquad
        \frac{e^{\tau\lambda}-e^{-\tau(q_{12}+q_{21})}}{\lambda+q_{12}+q_{21}},
    \end{equation*}
    with $\lambda\in\{\lambda_1^{\langle 1\rangle},\lambda_2^{\langle 1\rangle}\}$ (see the explicit formulas in Theorem~\ref{thm323}).
    
    It therefore suffices to ensure that $\lambda_1^{\langle 1\rangle}<0$ and $\lambda_2^{\langle 1\rangle}<0$. These numbers are exactly the eigenvalues of the constant matrix $\mathbf{P}_\alpha^{\langle 1\rangle}$ in Theorem~\ref{thm323}. If we take
    \begin{equation*}
        A:=\alpha(q_{12}+q_{21})-(\kappa_1+\kappa_2),
        \qquad
        B:=\kappa_1\kappa_2-\alpha\kappa_1q_{21}-\alpha\kappa_2q_{12},
    \end{equation*}
    the stated expressions can be rewritten as
    \begin{equation*}
        \lambda_1^{\langle 1\rangle}=\frac{-A+\sqrt{A^2-4B}}{2\alpha},
        \qquad
        \lambda_2^{\langle 1\rangle}=\frac{-A-\sqrt{A^2-4B}}{2\alpha}.
    \end{equation*}
    Under the parameters' admissibility for the limiting (unconditional) negative moment, we have $A>0$ and $B>0$, so $A^2-4B<A^2$ and hence $\sqrt{A^2-4B}<A$. This implies $\lambda_1^{\langle 1\rangle}<0$, while
    $\lambda_2^{\langle 1\rangle}<0$ is immediate from the formula.
    
    Consequently, as $\tau\to\infty$, we have $e^{\tau\lambda_1^{\langle 1\rangle}}\to0$ and $e^{\tau\lambda_2^{\langle 1\rangle}}\to0$, and also $e^{-\tau(q_{12}+q_{21})}\to0$.~Therefore,~$A_\alpha^{\langle 1\rangle}(\tau)\to0$ and $B_\alpha^{\langle 1\rangle}(\tau)\to0$. From the explicit forms of $A_\alpha^{\langle 0\rangle}(\tau)$ and $B_\alpha^{\langle 0\rangle}(\tau)$ in Theorem~\ref{thm323}, we~obtain
    \begin{equation*}
        \lim_{\tau\to\infty}A_\alpha^{\langle 0\rangle}(\tau)
        =\frac{-n_{11}}{\lambda_1^{\langle 1\rangle}}+\frac{-n_{12}}{\lambda_2^{\langle 1\rangle}},
        \qquad
        \lim_{\tau\to\infty}B_\alpha^{\langle 0\rangle}(\tau)
        =\frac{-n_{11}}{\lambda_1^{\langle 1\rangle}}+\frac{-n_{12}}{\lambda_2^{\langle 1\rangle}}.
    \end{equation*}
    Hence, the limit is the same for both regimes, and
    \begin{equation*}
        \lim_{\tau\to\infty}U_{\alpha,i}^{\langle -1\rangle}(\tau,R)
        =\frac{-n_{11}}{\lambda_1^{\langle 1\rangle}}+\frac{-n_{12}}{\lambda_2^{\langle 1\rangle}},
    \end{equation*}
    as claimed.
\end{proof}

\begin{corollary} \label{cor_uncon_2}
    According to Theorem~\ref{thm_uncon_2} and the system~\eqref{model4} with $\kappa_1=\kappa_2=\kappa, \theta_1=\theta_2=\theta$, and $\sigma_1=\sigma_2=\sigma$, we have
    \begin{equation*}
        \lim_{\tau\rightarrow\infty} U_{\alpha,i}^{\langle -1\rangle}(\tau,R) = \theta - \frac{\sigma^2}{2 \kappa} \left( \frac{1}{\alpha} + 1 \right).
    \end{equation*}
\end{corollary}
\begin{proof}
    Under $\kappa_1=\kappa_2=\kappa$, $\theta_1=\theta_2=\theta$ and $\sigma_1=\sigma_2=\sigma$,
    the coefficients in Theorem~\ref{thm_uncon_2} simplify and the limit becomes independent of $q_{12},q_{21}$.
    From Theorem~\ref{thm_uncon_2}, we have
    \[
    \lim_{\tau\to\infty}U_{\alpha,i}^{\langle -1\rangle}(\tau,R)
    =\frac{-n_{11}}{\lambda_1^{\langle 1\rangle}}+\frac{-n_{12}}{\lambda_2^{\langle 1\rangle}}.
    \]
    
    We first simplify the eigenvalues. Substituting $\kappa_1=\kappa_2=\kappa$ into the formulas in Theorem~\ref{thm_uncon_2}
    gives
    \[
    \lambda_1^{\langle 1\rangle}
    =\frac{1}{2\alpha}\Big(2\kappa-\alpha(q_{12}+q_{21})+\sqrt{\Delta}\Big),\qquad
    \lambda_2^{\langle 1\rangle}
    =\frac{1}{2\alpha}\Big(2\kappa-\alpha(q_{12}+q_{21})-\sqrt{\Delta}\Big),
    \]
    where
    \[
    \Delta=\big(-2\kappa+\alpha(q_{12}+q_{21})\big)^2-4\big(\kappa^2-\alpha\kappa(q_{12}+q_{21})\big)
    =\alpha^2(q_{12}+q_{21})^2.
    \]
    Hence $\sqrt{\Delta}=\alpha(q_{12}+q_{21})$, and therefore
    \[
    \lambda_1^{\langle 1\rangle}=\frac{\kappa}{\alpha},\qquad
    \lambda_2^{\langle 1\rangle}=\frac{\kappa}{\alpha}-(q_{12}+q_{21}),\qquad
    \lambda_2^{\langle 1\rangle}-\lambda_1^{\langle 1\rangle}=-(q_{12}+q_{21}).
    \]
    
    Next, under the same assumptions, we have
    $\overline{\gamma}_{\alpha,1}^{\langle0\rangle}=\overline{\gamma}_{\alpha,2}^{\langle0\rangle}
    =:\overline{\gamma}_{\alpha}^{\langle0\rangle}$, where
    \[
    \overline{\gamma}_{\alpha}^{\langle0\rangle}
    =\frac{1}{\alpha}\left(-\kappa\theta+\frac{1}{2}\sigma^2\left(\frac{1}{\alpha}+1\right)\right).
    \]
    Using the explicit expressions of $m_{ij}$ in Theorem~\ref{thm_uncon_2} and the eigenvalue simplifications above, one obtains
    $m_{11}=1$, $m_{12}=0$, $m_{21}=1$, and $m_{22}=0$; consequently
    \[
    n_{11}=\frac{\overline{\gamma}_{\alpha}^{\langle0\rangle}(m_{11}q_{21}+m_{21}q_{12})}{q_{12}+q_{21}}
    =\overline{\gamma}_{\alpha}^{\langle0\rangle},
    \qquad
    n_{12}=0.
    \]
    Substituting these into the limit formula yields
    \[
    \lim_{\tau\to\infty}U_{\alpha,i}^{\langle -1\rangle}(\tau,R)
    =\frac{-\overline{\gamma}_{\alpha}^{\langle0\rangle}}{\kappa/\alpha}
    =\frac{-1}{\kappa}\left(-\kappa\theta+\frac{1}{2}\sigma^2\left(\frac{1}{\alpha}+1\right)\right)
    =\theta-\frac{\sigma^2}{2\kappa}\left(\frac{1}{\alpha}+1\right),
    \]
    which proves the claim.
\end{proof}
Note that Corollaries~\ref{cor_uncon_1} and~\ref{cor_uncon_2} present results involving the parameters $\kappa$, $\theta$, and $\sigma$ for all values of $\alpha$ in their constraints, except when $\alpha = 1$ in Corollary~\ref{cor_uncon_1}. In this specific case, the NLD-CEV process reduces to the CIR process, and the unconditional moment depends only on $\theta$. This agrees with the results proposed in~\mbox{\cite{mongkolsin2025analytical,sutthimat2022closed}}.
\section{Euler--Maruyama method for SDEs with regime switching} \label{sec_exper}
\medskip

As detailed in Section~\ref{sec_CM}, our theoretical approach yield explicit formulas for conditional moments using the Euler--Maruyama (EM) method, as presented in Theorems~\ref{thm11} and~\ref{thm321}. An important consideration for market practitioners is the computational accuracy and efficiency of these newly derived formulas. To evaluate this, we compare results received from the formulas with those from the MC simulations. For implementing the hybrid system, we employ the EM method specifically adapted for SDEs with regime switching, as developed by Yuan and Mao~\cite{yuan2004convergence}, to simulate approximate solutions for SDEs with switching regimes. All simulations were performed using MATLAB R2021a on a system equipped with an Intel(R), Core(TM), i5-10500, CPU@3.10GHz, 16GB RAM, and Windows~11 Pro~(64-bit).
\subsection{Simulation algorithm}

Let $X_t$ represent a continuous-time, irreducible two-state Markov chain with the initial state $X_{t_0}$ and a step size $h >0$. Denote $Q$ as the generator matrix of $X_t$, where each transition rate from state $i$ to state $j$ is defined by $q_{ij} \geq 0$ and $q_{ii} = -\sum_{i \neq j} q_{ij}$. The probability of transitioning from state $i$ at time $t_0$ to state $j$ at time $t_0 + h$ is given by
\begin{align*}
    \mathbb{P}\big(X_{t_0+h} = j \mid X_{t_0} = i\big) =: P_{ij}(h) = \left(e^{hQ}\right)_{ij},
\end{align*}
for $i, j = 1, 2$. To simulate a discrete Markov chain $\{ X_k \}_{k \geq 0}$, where $X_{t_k} := X_k = X(kh)$, proceed as~follows:
\begin{enumerate}
    \item Compute the one-step transition probability matrix, $P(h) = e^{hQ}$.
    \item Set the initial state $X_0 = i_0$. Draw a uniformly distributed random number, $\omega \sim U[0,1]$, to determine the next state according to
    \begin{align*}
        X_k =
        \begin{cases}
            1, & \text{if } \omega \leq P_{i_{k}1}, \\
            2, & \text{if } P_{i_{k}1} < \omega \leq P_{i_{k}1} + P_{i_{k}2},
        \end{cases}
    \end{align*}
    and set $X_{k+1} = i_{k+1}$.
    \item Repeat Step 2 for each $k$ until $k^*$ is reached such that $h k^*$ equals the final time simulation, yielding the simulated trajectories of the process with the discrete Markov chain $\{ X_k \}_{k = 0}^{k^*}$.
\end{enumerate}
The EM method is then applied to the discrete Markov chain $\{ X_k \}_{k = 0}^{k^*}$ to simulate the processes in~\eqref{model3} and~\eqref{model4} and the approximate conditional moments. Let $\widehat{R}$ be the time-discretized approximation of $R$, with $\widehat{R}_k = \widehat{R}(t_k)$, $\widehat{R}_0 = R_{t_0}$, $\xi_0 = i_0$, and $h = t_{k+1} - t_k$. The EM approximation for~\eqref{model2} is given by
\begin{align}
    \widehat{R}_{k+1} = \widehat{R}_{k} + \kappa_{X_k} \Big( \theta_{X_k} \widehat{R}_k^{-(1-\beta)} - \widehat{R}_k \Big)h + \sigma_{X_k} \widehat{R}_k^\frac{\beta}{2} \sqrt{h} Z_k,
\end{align}
where $Z_k$ follows a standard normal distribution, as detailed in Algorithm~\ref{algSimu}.
\begin{algorithm}[H]
    \caption{MC simulation algorithm for the two-state continuous-time Markov chain with EM~discretization.} \label{algSimu}
    \floatname{algorithm}{Procedure}
    \renewcommand{\algorithmicrequire}{\textbf{Input:}}
    \renewcommand{\algorithmicensure}{\textbf{Output:}}
    \begin{algorithmic}[1]
        \Require Matrix $Q$ representing the infinitesimal generator, discretization step size $h$, terminal time horizon $T$, initial regime state $i_0$, initial process value $R_{t_0}$, model parameters: Mean-reversion rate $\kappa_{X_k}$, regime-dependent long-term means $\theta_{X_k}$, regime-dependent volatilities $\sigma_{X_k}$, elasticity parameter~$\beta$
        \Ensure Simulated trajectories of the regime process $\{ X_k, \, k = 0, 1, \dots, k^* \}$ and $\widehat{R}_k$
        \State Compute the transition probability matrix over the interval $h$ via the matrix exponential $P(h) = e^{hQ}$
        \State Initialize \ $t_0 = 0$ \ and \ $k = 0$
        \State Initialize \ $X_0 = i_0$ \ and \ $\widehat{R}_0 = R_{t_0}$
        \While {\ $t_k \leq T$}
            \State Generate uniform random variate $\omega \sim U[0,1]$
            \If { \ $\omega \leq P_{i_k 1}$}
                \State $X_{k+1} = 1$
            \Else
                \State $X_{k+1} = 2$
            \EndIf
            \State Compute \ $\widehat{R}_{k+1} = \widehat{R}_k + \kappa_{X_k} \left(\theta_{X_k} \widehat{R}_k^{-(1-\beta)} - \widehat{R}_k \right)h + \sigma_{X_k} \widehat{R}_k^{\frac{\beta}{2}} \sqrt{h} \, Z_k$
            \State Update \ $i_{k + 1} = X_{k+1}$, \ $t_{k+1} = t_k + h$, \ $k = k + 1$
        \EndWhile
        \State \Return \ Simulated trajectories of the regime process $\{ X_k, \, k = 0, 1, \dots, k^* \}$ and $\widehat{R}_k$
    \end{algorithmic}
\end{algorithm}

To simulate a sample path with 10,000~time steps, we configure the process in~\eqref{model3} by using the parameters $\kappa_1 = 0.01$, $\kappa_2 = 0.5$, $\theta_1 = 1$, and $\theta_2 = 0.5$ and the volatilities $\sigma_1 = 0.09$ and $\sigma_2 = 0.15$. For the process in~\eqref{model4}, the parameters are set to $\kappa_1 = -0.01$, $\kappa_2 = -0.5$, $\theta_1 = 1$, and $\theta_2 = 0.5$, with the volatilities $\sigma_1 = -0.09$ and $\sigma_2 = -0.15$. Both processes start from the initial conditions $X_0 = 1$, $R_0 = 1$, $\alpha = 0.5$, $q_{11} = -0.6$, $q_{12} = 0.6$, $q_{21} = 0.3$, and $q_{22} = -0.3$. In Figure 1(a),(b), red lines indicate intervals where the processes are in State~1, while blue lines represent intervals in State~2. The figures illustrate two scenarios: Over the time interval $t \in [0, 30]$, the process in~\eqref{model3} switches states 17 times, whereas the process in~\eqref{model4} switches 12 times.

As shown in Figure 2(a),(b), the results from MC simulations (depicted as colored circles) align perfectly with the predictions of Theorem~\ref{thm11} (depicted as solid lines) for each initial value $R_0 = 0.8, 1.2$ and the initial state $X_0 = 1, 2$. Similarly, Figure 2(c),(d) illustrates that the MC simulation's outcomes completely match those derived from Theorem~\ref{thm321}. However, a significant drawback of the MC simulations is the substantial computational time required to approximate the values for each initial $R_0$ and $X_0$. In contrast, our closed-form formulas provide exact solutions for all initial values and states with minimal computational effort, which we will explore further in the next section.
\newpage
\begin{figure}[H]
    \centering
    \begin{subfigure}{0.48\textwidth}
        \includegraphics[width=\textwidth]{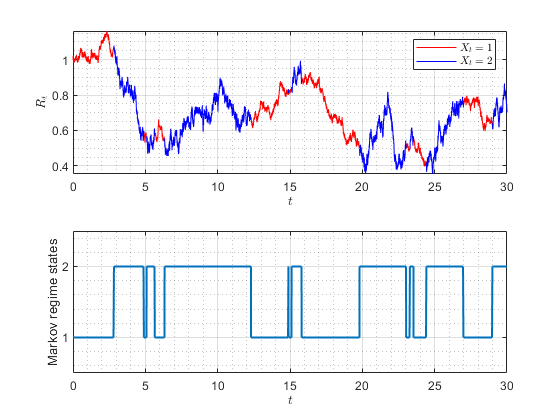}
        \caption{Regime-switching NLD-CEV process~\eqref{model3}}
        \label{1a}
    \end{subfigure}
    \hfill
    \begin{subfigure}{0.48\textwidth}
        \includegraphics[width=\textwidth]{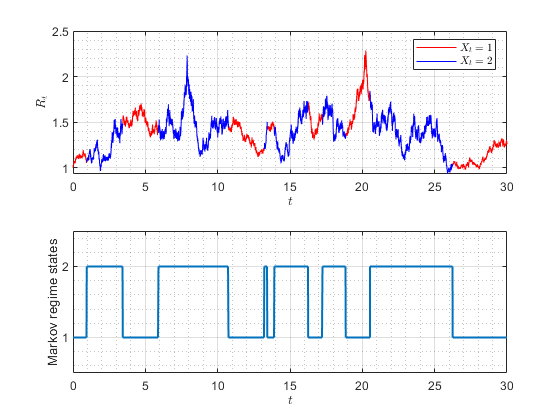}
        \caption{Regime-switching NLD-CEV process~\eqref{model4}}
        \label{1b}
    \end{subfigure}
    \caption{Sample paths of the two-state regime-switching NLD-CEV process using the EM method for SDEs.}
\label{fig1}
\end{figure}


\begin{figure}[H]
    \centering
    \begin{subfigure}{0.496\textwidth}
        \includegraphics[width=\textwidth]{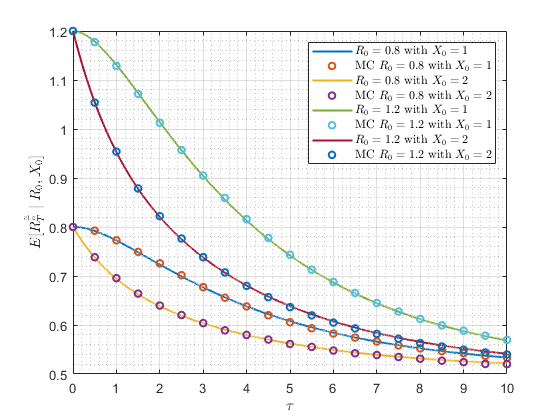}
        \caption{The first conditional moments of the regime-switching CIR process ($\beta=1$) when $n=1$ and $\alpha=1$.}
        \label{2a}
    \end{subfigure}
    \hfill
        \begin{subfigure}{0.496\textwidth}
        \includegraphics[width=\textwidth]{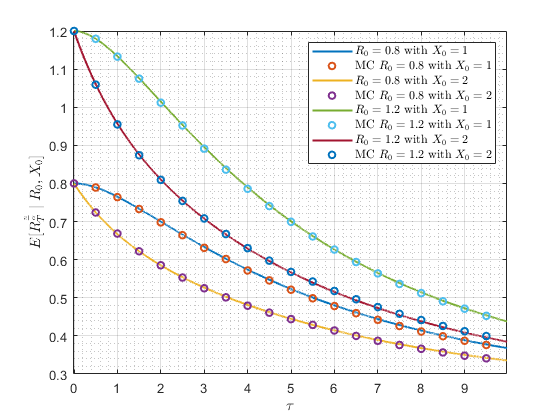}
        \caption{The first conditional moments of the regime-switching OU process ($\beta=0$) when $n=2$ and $\alpha=2$.}
        \label{2b}
    \end{subfigure}
        \begin{subfigure}{0.496\textwidth}
        \includegraphics[width=\textwidth]{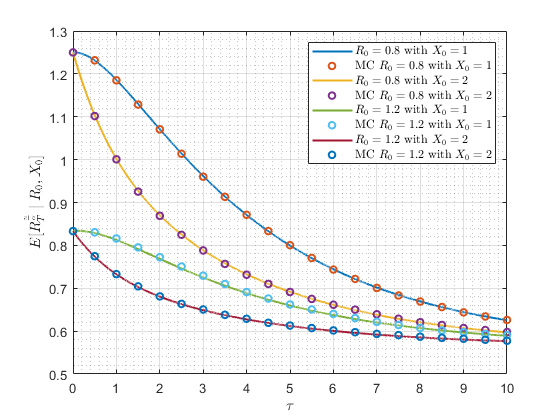}
        \caption{The first inverse moments of the regime-switching IF \\process ($\beta=3$) when $n=-1$ and $\alpha=1$.}
        \label{2c}
    \end{subfigure}
            \begin{subfigure}{0.496\textwidth}
        \includegraphics[width=\textwidth]{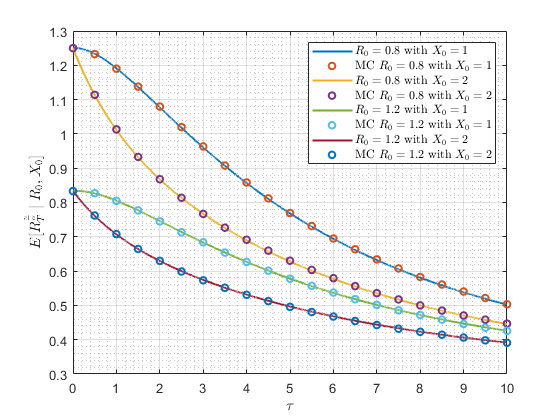}
        \caption{The first inverse moments when $\beta=\frac{5}{2}$, $n=-2$, and $\alpha=2$.}
        \label{2d}
    \end{subfigure}
    \caption{Conditional moments of the regime-switching NLD-CEV process.}
\label{fig2}
\end{figure}


\subsection{Accuracy and efficiency}
\medskip
Although, we have derived the exact solutions presented in Section~\ref{sec_CM}, MC simulations are still necessary to confirm the correctness of our results and to assure practitioners that our findings are applicable to real-world problems. Additionally, it is important to assess the computational effort involved. To evaluate the accuracy and efficiency of our results $U_{\alpha,i}^{\langle n \rangle}(\tau, R)$ proposed in Theorems~\ref{thm11} and~\ref{thm321}, compared with the MC simulations $U_{\alpha,i}^{\langle n, M \rangle}(\tau, R)$, we define the absolute relative errors
\begin{equation*}
    E_{U_{\alpha,i}}^{\langle \gamma \rangle}(\tau,R) := \left| \frac{U_{\alpha,i}^{\langle n\rangle}(\tau,R) - U_{\alpha,i}^{\langle n, M \rangle}(\tau,R)}{U_{\alpha,i}^{\langle n\rangle}(\tau,R)} \right|,
\end{equation*}
for all $(\tau,R)\in [0,\infty)\times(0,\infty)$ and $i \in \mathcal{M}_2$. As displayed in Figure~\ref{fig2}, we demonstrate the results for the~CIR process in~\eqref{model3} with $\alpha = 1$, and the IF process in~\eqref{model4} with $\alpha = 1$.

Table~\ref{tab1} shows the comparison between our exact closed-form solutions for $\mathbf{E}^{\mathbb{Q}}\big[ R_T^{{n}/{\alpha}} ~|~ R_t= R, X_t = i \big]$ and the approximations obtained from MC simulations. The results indicate that as the number of simulation paths increases, the MC simulations converge towards our exact values, with average relative errors decreasing from the order of $10^{-3}$ to $10^{-4}$ or less. Moreover, our closed-form formulas significantly outperform the MC simulations in terms of computational efficiency. While the running time for the MC simulations escalates dramatically with more paths (ranging from approximately 6~sec to over~266~sec), our exact solutions are computed in less than 0.01~seconds consistently, regardless of the initial conditions or parameters. This highlights the superiority of our method in providing exact and computationally efficient solutions compared with the approximate and resource-intensive MC~simulations.
\begin{table}[H]
    \caption{Comparison of the average relative errors between our results with MC simulations for $\mathbf{E}^{\mathbb{Q}}\big[ R_T^{{n}/{\alpha}} ~|~ R_t= R, X_t = i \big]$.} \label{tab1}
    \small
    \begin{tabular}{llllllllllll}
    \toprule
        Number &
          \multicolumn{5}{l}{Theorem~\ref{thm11} with $X_0 = 1$, $\alpha = 1$, and $n = 1$} &
           &
          \multicolumn{5}{l}{Theorem~\ref{thm11} with $X_0 = 2$, $\alpha = 1$, and $n = 1$} \\ \cmidrule(lr){2-6} \cmidrule(l){8-12} 
        of paths &
          \multicolumn{2}{l}{Average relative error} &
           &
          \multicolumn{2}{l}{Average running time} &
           &
          \multicolumn{2}{l}{Average relative error} &
           &
          \multicolumn{2}{l}{Average running time} \\ \cmidrule(lr){2-3} \cmidrule(lr){5-6} \cmidrule(lr){8-9} \cmidrule(l){11-12} 
         &
          $R = 0.8$ &
          $R = 1.2$ &
           &
          MC &
          This work &
           &
          $R = 0.8$ &
          $R = 1.2$ &
           &
          MC &
          This work \\ \cmidrule(r){1-3} \cmidrule(lr){5-6} \cmidrule(lr){8-9} \cmidrule(l){11-12} 
        { }\,5,000 &
          1.456e-03 &
          1.765e-03 &
           &
          6.243 &
          \multirow{5}{*}{0.00925} &
           &
          3.784e-03 &
          9.716e-04 &
           &
          8.870 &
          \multirow{5}{*}{0.00451} \\
        10,000 &
          1.326e-03 &
          9.037e-04 &
           &
          18.389 &
           &
           &
          2.184e-03 &
          7.492e-04 &
           &
          26.268 &
           \\
        20,000 &
          1.307e-03 &
          5.580e-04 &
           &
          47.355 &
           &
           &
          1.124e-03 &
          5.715e-04 &
           &
          65.713 &
           \\
        40,000 &
          4.262e-04 &
          4.576e-04 &
           &
          118.399 &
           &
           &
          4.017e-04 &
          5.477e-04 &
           &
          147.773 &
           \\
        80,000 &
          2.476e-04 &
          3.042e-04 &
           &
          266.572 &
           &
           &
          1.971e-04 &
          1.282e-04 &
           &
          297.013 &
        \\ \midrule
        Number &
          \multicolumn{5}{l}{Theorem~\ref{thm321} with $X_0 = 1$, $\alpha = 1$, and $n = -1$} &
           &
          \multicolumn{5}{l}{Theorem~\ref{thm321} with $X_0 = 2$, $\alpha = 1$, and $n = -1$} \\ \cmidrule(lr){2-6} \cmidrule(l){8-12} 
        of paths &
          \multicolumn{2}{l}{Average relative error} &
           &
          \multicolumn{2}{l}{Average running time} &
           &
          \multicolumn{2}{l}{Average relative error} &
           &
          \multicolumn{2}{l}{Average running time} \\ \cmidrule(lr){2-3} \cmidrule(lr){5-6} \cmidrule(lr){8-9} \cmidrule(l){11-12} 
         &
          $R = 0.8$ &
          $R = 1.2$ &
           &
          MC &
          This work &
           &
          $R = 0.8$ &
          $R = 1.2$ &
           &
          MC &
          This work \\ \cmidrule(r){1-3} \cmidrule(lr){5-6} \cmidrule(lr){8-9} \cmidrule(l){11-12} 
        { }\,5,000 &
          1.256e-03 &
          1.465e-03 &
           &
          5.372 &
          \multirow{5}{*}{0.00861} &
           &
          2.279e-03 &
          9.149e-04 &
           &
          7.154 &
          \multirow{5}{*}{0.00422} \\
        10,000 &
          1.011e-03 &
          8.173e-04 &
           &
          16.622 &
           &
           &
          2.001e-03 &
          6.866e-04 &
           &
          24.457 &
           \\
        20,000 &
          8.782e-04 &
          5.473e-04 &
           &
          44.235 &
           &
           &
          1.021e-03 &
          4.738e-04 &
           &
          61.203 &
           \\
        40,000 &
          4.268e-04 &
          4.375e-04 &
           &
          116.781 &
           &
           &
          3.657e-04 &
          5.249e-04 &
           &
          142.817 &
           \\
        80,000 &
          1.945e-04 &
          1.657e-04 &
           &
          258.671 &
           &
           &
          1.854e-04 &
          1.057e-04 &
           &
          288.701 &
        \\ \bottomrule
    \end{tabular}
\end{table}

\section{Financial applications on the VIX} \label{sec_application}
\medskip
The VIX, a 30-day implied volatility measure derived from the Standard \& Poor’s (S\&P) 500 index options, is essential for assessing market sentiment and managing volatility risk. Accurate pricing of VIX derivatives depends on using a model that captures the complex nature of volatility. This study presents a novel approach to VIX option pricing, introducing a closed-form formula based on the NLD-CEV process with regime-switching. The NLD-CEV process offers several advantages over traditional models, such as the CIR model, by incorporating mean reversion and enabling data-driven estimation of the elasticity parameter~$\beta$. This flexibility addresses a key limitation of commonly used models like the square root process and $3/2$-SVM, which assume fixed values of $\beta$ at 1 and 3, respectively. Empirical evidence suggests that $\beta$ often falls outside this range, underscoring the need for a model that allows for data-driven parameter estimation \cite{dupoyet2011simplified, tong2017modelling}.

In this section, the numerical illustrations focus on theoretical development and illustrative numerical experiments rather than calibration to market data. Since VIX data are not available within this project, we do not calibrate or estimate model parameters from VIX time series; hence, the parameter choices used in the MC experiments are not presented as statistical estimates of~VIX dynamics. Instead, we use suitable parameter choices that satisfy the model's assumptions and are intended to demonstrate the implications of regime-switching dynamics and our closed-form formulas in a transparent manner.

To support this simulation design, we follow the methodological discussion in the regime-switching pricing literature. It is common practice to evaluate pricing models using parameters obtained for the full sample; alternatively, Driffill et al.~\cite{driffill2009effects} considered a real-time pricing approach in which the prices were computed using the best available estimates of the parameters at each point in time, based on recursively estimating the model with a sequentially enlarged sample. Although such recursive estimation is not implemented here, the main implication is that the results may depend on the choice of the parameters that are allowed to switch across regimes, and different parameterizations (e.g., regime-switching in volatility only versus allowing multiple parameters to switch) can lead to different pricing performance~\cite{driffill2009effects}. Motivated by this point, we complement the baseline parameter choices by reporting a sensitivity analysis under alternative parameter settings and by illustrating how the results change under different regime-switching parameterizations.

Leveraging the theoretical basis of the NLD-CEV process, this research develops a VIX futures and options pricing formula with practical implications. The ability to derive a closed-form solution for fractional-order conditional moments in the NLD-CEV process enables efficient and accurate option price calculations. This methodology bridges the gap between advanced mathematical modeling and~practical financial applications, significantly enhancing option pricing techniques for volatility~derivatives.


\subsection{Futures pricing strategy: Offsetting positions}

The strategy of buying/selling futures contracts to close an existing short/long position before expiration is commonly referred to as offsetting or closing out the position. In futures trading, entering an opposite trade to neutralize the existing position is standard practice. For example, if we have a long position in a futures contract (we have agreed to buy the underlying asset at a future date), and we want to exit the position before expiration, we can go short to close long. This means the following:
\begin{enumerate}
    \item Selling an equivalent futures contract to offset our long position;
    \item The short position cancels out the long position, effectively closing our obligation to buy the asset;
    \item We can avoid the need to take physical delivery of the underlying asset upon expiration by closing the~position.
\end{enumerate}

In practice, this strategy can avoid physical delivery and benefit to many traders speculating on price movements who do not wish to handle the actual asset. Additionally, the strategy can be used to lock in profits or limit losses because closing the position allows traders to realize gains or prevent further~losses.

In this section, we introduce two trading strategies based on our closed-form solutions: The short-to-close long position and the long-to-close short position. These strategies leverage our analytical results to optimize the timing of closing positions in a regime-switching market. The steps for implementing these strategies are outlined as follows.
\begin{enumerate}
    \item Determine the current price $R_t$ of the underlying asset and identify the prevailing regime state $X_t$ in the market. This involves analyzing market indicators and utilizing regime-switching models to accurately ascertain the state of the market.
    \item On the basis of the market analysis, decide whether to open a long or short position on the asset. This decision should be informed by expectations of future price movements and the identified market~regime.
    \item Apply our closed-form solutions for the conditional expectations $\mathbf{E}^{\mathbb{Q}}\big[ R_T^{{n}/{\alpha}} ~\big|~ R_t= R, X_t = i \big]$, as presented in Theorems~\ref{thm11} and~\ref{thm321}, to determine the optimal time to close the position.
    \begin{itemize}
        \item For a long position: Implement the short-to-close strategy. Use the closed-form solution to forecast the expected future price dynamics under different regimes. This enables the identification of the optimal exit point to sell the asset and close the long position, maximizing profit or minimizing loss.
        \item For a short position: Implement the long-to-close strategy. Use the closed-form solution to predict the asset\rq s price trajectory. This assists in determining the optimal time to buy the asset and close the short position, ensuring the most favorable financial outcome.
    \end{itemize}
    \item On the basis of the insights from the closed-form solutions, execute the trade to close the position at the calculated optimal time. This ensures that the trading strategy is aligned with the analytical predictions, enhancing the effectiveness of the approach.
\end{enumerate}
By following these steps, traders can effectively apply our analytical results to real-world scenarios. The use of exact closed-form formulas provides a significant advantage over numerical methods (see also~\cite{li2016trading}), offering precise and computationally efficient solutions. This approach allows practitioners to optimize their trading strategies in markets characterized by regime-switching behavior, ultimately improving decision-making and financial performance.

\begin{figure}[!ht]
    \centering
    \begin{subfigure}[!ht]{0.495\textwidth}
        \includegraphics[width=\textwidth]{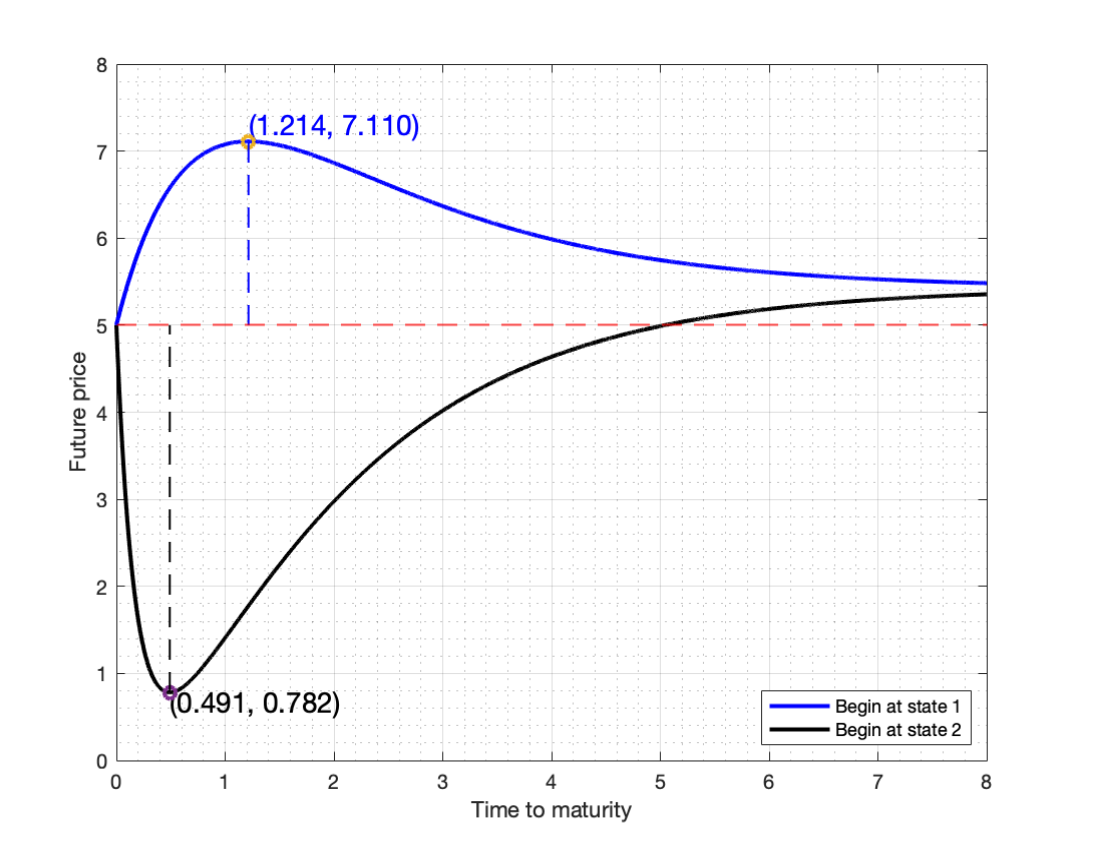}
        \caption{Given $Q_{1}$: $q_{11} = -0.3$, $q_{12} = 0.3$, $q_{21} = 0.3$, $q_{22} = -0.3$.}
        \label{3a}
    \end{subfigure}
    \begin{subfigure}[!ht]{0.495\textwidth}
        \includegraphics[width=\textwidth]{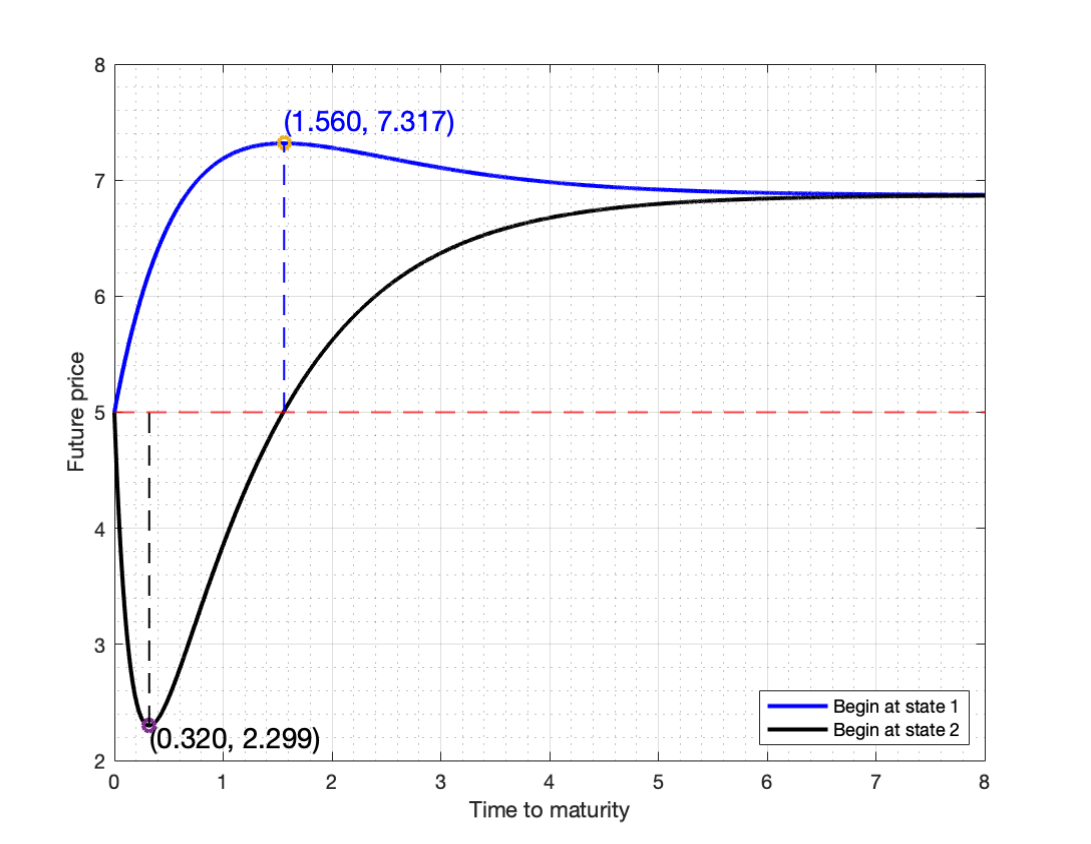}
        \caption{Given $Q_{2}$: $q_{11} = -0.3$, $q_{12} = 0.3$, $q_{21} = 0.7$, $q_{22} = -0.7$.}
        \label{3b}
    \end{subfigure}
    \caption{Future prices for the CIR process with the two-state regime-switching.}
\label{fig3}
\end{figure}

To provide a concrete example of seeking the optimal time based on our results, we consider Theorem~\ref{thm313} with $\alpha = 1$ and $n = 1$, which corresponds to the conditional first moment of the CIR process. We use the following parameters: $R = 5$, $\kappa_1 = 1$, $\kappa_2 = 3$, $\theta_1 = 10$, $\theta_2 = 2$, $\sigma_1 = 0.1$, and~$\sigma_2 = 0.2$. As displayed in Figure~\ref{fig3}, the blue line represents the futures prices starting from $X_t = 1$, the black line represents those starting~from~$X_t = 2$, and the red dashed line marks the initial price~$R = 5$. From Figure 3(a), using our formula, we conclude that when the current state is $X_t = 1$, the optimal strategy is to short the futures contract and close the position (short-to-close strategy) at time to maturity $\tau = 1.214$, resulting in a profit of $7.110 - 5 = 2.110$ units per contract. Conversely, when the current state is $X_t = 2$, the optimal strategy is to long the futures contract and close the position~(long-to-close strategy) at time to maturity $\tau = 0.491$, resulting in a profit of $5 - 0.782 = 4.218$ units per contract. Similarly, the return can be analyzed and calculated (maximizing profit or minimizing loss) in the same way, as shown in Figure 3(b).

Additionally, the transition rates for the generator matrix of Figure 3(a) are equal and symmetric between the two states, and the stationary distribution is $\pi = (0.5, 0.5)$. It implies that the system is balanced and that both states occur equally often in the long run. In the case of Figure 3(b), the rate from State~2 to State~1 (0.7) is higher than that from State~1 to State~2 (0.3), which means that State~2 is less stable and the  system leaves State~2 more quickly. The stationary distribution is $\pi = (0.7, 0.3)$. It means that State~1 occurs more often (70\% of the time) and State~2 occurs less often (30\% of the time). In summary, the system prefers State~1 due to the higher exit rate from State~2 and the likelihood of being in State~1 increases, making it the dominant state. As see in Figure 3(b), it can be noted that the black line is pulled back to the long-run term, which can be calculated by using Theorem~\ref{thm_uncon_1}.

In the context of real-world implications, the differences between the generator matrices $Q_1$ and $Q_2$ have significant implications for modeling and applications across various fields. In $Q_1$, the symmetric transition rates ($q_{12} = q_{21} = 0.3$) indicate that both states occur equally often, modeling systems where the conditions alternate regularly without a preference for either state. This symmetry suggests a balanced system, such as an economy with regular cycles between growth and recession. In contrast, $Q_2$ exhibits asymmetric transition rates, with a higher rate of transitioning from State~2 to State~1 ($q_{21} = 0.7$, $q_{12} = 0.3$), resulting in State~1 occurring more frequently. This asymmetry models systems, where one state is more stable or dominant, reflecting scenarios like prolonged economic expansions interrupted by brief recessions. Understanding these dynamics allows for more accurate modeling of real-world systems, enabling better predictions and strategic planning. By adjusting the transition rates in the generator matrix, we can simulate different scenarios and assess their impact on the system's behavior, which is crucial for risk management and decision-making processes in finance, economics, engineering, and healthcare.

It is important to emphasize that the trading strategies described above are developed under a frictionless market assumption. In practical implementations, transaction costs, bid--ask spreads, and other market frictions may directly affect both the execution and profitability of these timing-based strategies. Since the closing time is identified from regime-dependent conditional expectation curves derived in closed form, changes in the market regimes may induce frequent position adjustments or re-optimization of exit timing. Such adjustments can accumulate substantial trading costs, which may significantly reduce the net returns or even offset the theoretical gains implied by the analytical results. Moreover, liquidity constraints and execution delays may further affect the feasibility and effectiveness of implementing the proposed strategies in real-market conditions. Accordingly, the numerical illustrations presented in this section should be interpreted as demonstrating the theoretical potential of the approach rather than guaranteed performance in practice.


\subsection{VIX options pricing}
\medskip

This section applies the derived closed-form formula to price VIX options within a two-state regime-switching NLD-CEV process. The focus is on European call options on the VIX index, a critical tool in volatility trading and risk management. Consider a European call option with the maturity date $T$ and the strike price $K$. The fair value of this option at any time $t \in [0, T]$ is defined by
\begin{equation} \label{vix_option_price}
    C_T(t, R, K, i) = e^{-r^*\tau} \mathbf{E}^\mathbb{Q}\big[ \max(R_T - K, 0) \mid R_t = R, X_t = i \big],
\end{equation}
where $r^*$ is the risk-free rate, $\tau = T - t$ denotes the time to maturity, $R_t$ represents the VIX level at time $t$, and $X_t$ indicates the regime state at time $t$. To efficiently compute the expectation in~\eqref{vix_option_price}, the Laguerre expansion method by Dufresne~\cite{Dufrense2000option} is used. This approach expresses the European call option price as an infinite series,
\begin{equation} \label{laguerre_expansion}
    C_T(t, R, K, i) = e^{-r^*\tau} K^b e^{-K} \sum_{j=0}^{\infty} c_j(t, R, T, i) \, \mathbf{L}_j^a(K),
\end{equation}
where $a, b \in \mathbb{R}$ satisfy $a > 2\max(b, 0) - 1$, and $\mathbf{L}_j^{a}(\cdot)$ represents the generalized Laguerre polynomial~\cite{Dufrense2000option} of order $j$ with the parameter $a > -1$. The coefficients $c_j$ are given in terms of the conditional moments of $R_T$
\begin{equation} \label{laguerre_coefficients}
    c_j(t, R, T, i) = \sum_{n=0}^j \frac{(-1)^n \ \mathbf{E}^\mathbb{Q}\Big[R_T^{\frac{\gamma_n}{\alpha}} \mid R_t = r, X_t = i\Big]}{\frac{\gamma_n}{\alpha} \big(\frac{\gamma_n}{\alpha} - 1\big) \Gamma(n + a + 1)},
\end{equation}
for $j \in \mathbb{N}_0$, where $\frac{\gamma_n}{\alpha} = a - b + n + 2$. Implementing this option pricing formula involves several steps. First, suitable values for $a$ and $b$ are chosen to satisfy $a > 2\max(b, 0) - 1$ and to ensure that $\frac{\gamma_n}{\alpha} \in \mathbb{N}$ for all~$n \in [0, j]$. This choice of parameters guarantees the convergence of the Laguerre series and facilitates calculation of the coefficients. Next, the closed-form formula for the NLD-CEV process expectation is used to obtain the moments $\mathbf{E}^\mathbb{Q}\big[ R_T^{{\gamma_n}/{\alpha}} \mid R_t = R, X_t = i \big]$. These moments are subsequently applied to calculate the coefficients $c_j$ in~\eqref{laguerre_coefficients}. Finally, the option price is computed using a truncated version of the series in~\eqref{laguerre_expansion}.

This method offers practical advantages for VIX option pricing, notably computational efficiency and analytical tractability through closed-form expressions. Within the NLD-CEV modeling perspective, the resulting pricing formulas provide a flexible mechanism to capture the key qualitative features of VIX dynamics under suitable parameter choices. While the present study focuses on theoretical development and illustrative numerical experiments rather than calibration to market data, the results suggest that the approach can serve as a useful building block for volatility pricing and hedging. Further empirical calibration and validation using historical VIX and VIX market data would be a natural next step to assess the pricing performance and practical risk management impact. Overall, this work contributes tractable option pricing methodology that connects rigorous stochastic modeling with applications in quantitative finance.


\section{Conclusions} \label{sec_conclusion}

In this study, we presented an analytical framework for the fractional-order conditional moments of the NLD-CEV process with regime-switching dynamics. By integrating a nonlinear drift component and accommodating elasticity in the variance, our approach offers a versatile modeling tool for financial markets, especially in pricing derivatives like VIX futures, where volatility and market shifts play critical roles. The hybrid system approach adopted here allows for exact closed-form solutions, which significantly enhances computational efficiency compared with traditional MC simulations.

Our numerical results underscore the computational advantages of the closed-form formulas derived in this study. In particular, we observe substantial reductions in the computational time, with an accuracy that aligns closely with the MC simulation's outcomes. This computational efficiency is invaluable in practical applications where rapid pricing and risk assessment are essential. Furthermore, our results demonstrate that the NLD-CEV process captures nuanced dynamics by adjusting to changes in the economic conditions through regime-switching, thereby providing more accurate reflections of market realities than many traditional models.

The applications explored here, specifically for VIX futures and options pricing, highlight the practical utility of our model in financial engineering. The model’s adaptability to shifts in market sentiment and volatility makes it well-suited for environments characterized by sudden economic changes. By incorporating regime-switching elements, the NLD-CEV process not only enhances accuracy in volatility modeling but also addresses some limitations inherent in simpler processes by capturing nonlinear responses to market shocks. More broadly, the proposed framework may be applicable beyond volatility markets, including interest rate modeling and credit risk or credit derivative pricing, where regime-switching dynamics are commonly used to capture structural changes, policy shifts, or distress events.

While the present analysis focuses on a two-state Markov chain for analytical clarity, the proposed framework is not inherently restricted to this setting. In principle, the hybrid PDE system may be extended to an arbitrary finite number of regimes ($m>2$), resulting in a system of $m$ coupled equations that captures more complex market environments, albeit with increased analytical and computational complexity. Moreover, although an irreducible Markov chain is assumed to ensure well-defined regime dynamics, the methodology may be adapted to non-irreducible or partially absorbing chains under appropriate modeling considerations, which are particularly relevant in applications such as credit risk or distress modeling.

Future research could extend this work by exploring multi-regime settings and incorporating additional stochastic elements to further refine the model’s adaptability to market complexities. In addition, it is important to note that Assumptions~1--6 are imposed primarily to ensure the existence and uniqueness of the solution and to facilitate analytical tractability. 
While these assumptions enable closed-form derivations, they may be restrictive in practical settings where the model parameters exhibit abrupt changes rather than smooth temporal variation. 
An interesting direction for future research is to relax these assumptions by considering weaker specifications, such as piecewise constant or regime-wise parameter dynamics, which would require additional conditions to preserve the well-posedness. Empirical validation using a broader dataset of financial instruments could provide insights into optimizing the model's parameters for specific financial contexts. In particular, calibrating the proposed model to real VIX derivative transaction data and conducting comparative pricing analysis against established benchmark models would offer a more comprehensive assessment of its real-world pricing accuracy and practical performance. This research contributes a significant step toward more efficient and accurate pricing methodologies, bridging advanced stochastic modeling techniques with concrete financial applications.


\begin{thebibliography}{1}

\bibitem{ahn1999parametric}
D.~H. Ahn and B.~Gao.
\newblock A parametric nonlinear model of term structure dynamics.
\newblock {\em The Review of Financial Studies}, 12(4):721--762, 1999.

\bibitem{anderson2012continuous}
W.~J. Anderson.
\newblock {\em Continuous-time Markov chains: An applications-oriented approach}.
\newblock Springer Science \& Business Media, New York, 1st edition, 2012.

\bibitem{araneda2021sub}
A.~A. Araneda and N.~Bertschinger.
\newblock The sub-fractional CEV model.
\newblock {\em Physica A: Statistical Mechanics and its Applications}, 573:125974, 2021.

\bibitem{baran2013feynman}
N.~A. Baran, G.~Yin, and C.~Zhu.
\newblock Feynman--Kac formula for switching diffusions: connections of systems
of partial differential equations and stochastic differential equations.
\newblock {\em Advances in Difference Equations}, 2013:315, 2013.

\bibitem{boyarchenko2009american}
S.~Boyarchenko and S.~Levendorskii.
\newblock American options in regime-switching models.
\newblock {\em SIAM Journal on Control and Optimization}, 48:1353--1376, 2009.

\bibitem{buffington2002regime}
J.~Buffington and R.~J. Elliott.
\newblock Regime switching and European options.
\newblock In {\em Stochastic Theory and Control}, Lecture Notes in Control and
Information Sciences. Springer, Berlin, 2002.

\bibitem{cao2021pricing}
J.~Cao, J.-H. Kim, and W.~Zhang.
\newblock Pricing variance swaps under hybrid CEV and stochastic volatility.
\newblock {\em Journal of Computational and Applied Mathematics}, 386:113220, 2021.

\bibitem{carr2017leverage}
P.~Carr and L.~Wu.
\newblock Leverage effect, volatility feedback, and self-exciting market disruptions.
\newblock {\em Journal of Financial and Quantitative Analysis}, 52:2119--2156, 2017.

\bibitem{chan1992empirical}
K.~C. Chan, G.~A. Karolyi, F.~A. Longstaff, and A.~B. Sanders.
\newblock An empirical comparison of alternative models of the short-term interest rate.
\newblock {\em The Journal of Finance}, 47:1209--1227, 1992.

\bibitem{chumpong2024closed}
K.~Chumpong, K.~Mekchay, F.~Nualsri, and P.~Sutthimat.
\newblock Closed-form formula for the conditional moment-generating function under a
regime-switching, nonlinear drift CEV process, with applications to option pricing.
\newblock {\em Mathematics}, 12:2667, 2024.

\bibitem{chumpong2022simple}
K.~Chumpong, R.~Tanadkithirun, and C.~Tantiwattanapaibul.
\newblock Simple closed-form formulas for conditional moments of inhomogeneous nonlinear drift
constant elasticity of variance process.
\newblock {\em Symmetry}, 14:1345, 2022.

\bibitem{cox1975notes}
J.~Cox.
\newblock Notes on option pricing I: Constant elasticity of variance diffusions.
\newblock {\em Unpublished note, Stanford University, Graduate School of Business}, 1975.

\bibitem{cox1976valuation}
J.~C. Cox and S.~A. Ross.
\newblock The valuation of options for alternative stochastic processes.
\newblock {\em Journal of Financial Economics}, 3:145--166, 1976.

\bibitem{driffill2009effects}
J.~Driffill, T.~Kenc, M.~Sola, and F.~Spagnolo.
\newblock The effects of different parameterizations of Markov-switching in a CIR model of bond pricing.
\newblock {\em Studies in Nonlinear Dynamics \& Econometrics}, 13, 2009.

\bibitem{Dufrense2000option}
D.~Dufresne.
\newblock Laguerre series for Asian and other options.
\newblock {\em Mathematical Finance}, 10(4):407--428, 2000.

\bibitem{dupoyet2011simplified}
B.~Dupoyet, R.~T. Daigler, and Z.~Chen.
\newblock A simplified pricing model for volatility futures.
\newblock {\em Journal of Futures Markets}, 31(4):307--339, 2011.

\bibitem{dumas1998implied}
B.~Dumas, J.~Fleming, and R.~E. Whaley.
\newblock Implied volatility functions: Empirical tests.
\newblock {\em The Journal of Finance}, 53:2059--2106, 1998.

\bibitem{elliott2013pricing}
R.~J. Elliott and G.~H. Lian.
\newblock Pricing variance and volatility swaps in a stochastic volatility model with regime switching:
Discrete observations case.
\newblock {\em Quantitative Finance}, 13:687--698, 2013.

\bibitem{grzelak2011heston}
L.~A. Grzelak and C.~W. Oosterlee.
\newblock On the Heston model with stochastic interest rates.
\newblock {\em SIAM Journal on Financial Mathematics}, 2(1):255--286, 2011.

\bibitem{haddad2021repeated}
R.~E. Haddad.
\newblock Repeated integration and explicit formula for the $n$-th integral of $x^m (\ln x)^{m'}$.
\newblock {\em arXiv preprint arXiv:2102.11723}, 2021.

\bibitem{he2023new}
X.~J. He and S.~Lin.
\newblock A new nonlinear stochastic volatility model with regime switching stochastic mean reversion
and its applications to option pricing.
\newblock {\em Expert Systems with Applications}, 212:118742, 2023.

\bibitem{hull1990pricing}
J.~Hull and A.~White.
\newblock Pricing interest-rate-derivative securities.
\newblock {\em The Review of Financial Studies}, 3:573--592, 1990.

\bibitem{jones2003nonlinear}
C.~S. Jones.
\newblock Nonlinear mean reversion in the short-term interest rate.
\newblock {\em The Review of Financial Studies}, 16:793--843, 2003.

\bibitem{li2016trading}
J.~Li.
\newblock Trading VIX futures under mean reversion with regime switching.
\newblock {\em International Journal of Financial Engineering}, 3:1650021, 2016.

\bibitem{lin2020regime}
S.~Lin and X.~J. He.
\newblock A regime switching fractional Black--Scholes model and European option pricing.
\newblock {\em Communications in Nonlinear Science and Numerical Simulation}, 85:105222, 2020.

\bibitem{liu2024solving}
T.~Liu, F.~Soleymani, and M.~Z. Ullah.
\newblock Solving multi-dimensional European option pricing problems by integrals of the inverse quadratic
radial basis function on non-uniform meshes.
\newblock {\em Chaos, Solitons \& Fractals}, 185:115156, 2024.

\bibitem{marsh1983stochastic}
T.~A. Marsh and E.~R. Rosenfeld.
\newblock Stochastic processes for interest rates and equilibrium bond prices.
\newblock {\em The Journal of Finance}, 38(2):635--646, 1983.

\bibitem{mirevski2010some}
S.~Mirevski and L.~Boyadjiev.
\newblock On some fractional generalizations of the Laguerre polynomials and the Kummer function.
\newblock {\em Computers \& Mathematics with Applications}, 59(3):1271--1277, 2010.

\bibitem{mongkolsin2025analytical}
M.~Mongkolsin, K.~Mekchay, and P.~Sutthimat.
\newblock Analytical solutions for time-fractional Cauchy problem based on OU, CIR and Jacobi processes
with time-dependent parameters.
\newblock {\em Results in Applied Mathematics}, 28:100657, 2025.

\bibitem{rujivan2014simple}
S.~Rujivan and S.-P. Zhu.
\newblock A simple closed-form formula for pricing discretely-sampled variance swaps under the Heston model.
\newblock {\em The ANZIAM Journal}, 56(1):1--27, 2014.

\bibitem{rujivan2023analytically}
S.~Rujivan, A.~Sutchada, K.~Chumpong, and N.~Rujeerapaiboon.
\newblock Analytically computing the moments of a conic combination of independent noncentral chi-square
random variables and its application for the extended Cox--Ingersoll--Ross process with time-varying dimension.
\newblock {\em Mathematics}, 11(5):1276, 2023.

\bibitem{shen2013pricing}
Y.~Shen and T.~K. Siu.
\newblock Pricing variance swaps under a stochastic interest rate and volatility model with regime-switching.
\newblock {\em Operations Research Letters}, 41:180--187, 2013.

\bibitem{sutthimat20222closed}
P.~Sutthimat, K.~Mekchay, and S.~Rujivan.
\newblock Closed-form formula for conditional moments of generalized nonlinear drift CEV process.
\newblock {\em Applied Mathematics and Computation}, 428:127213, 2022.

\bibitem{sutthimat2022closed}
P.~Sutthimat and K.~Mekchay.
\newblock Closed-form formulas for conditional moments of inhomogeneous Pearson diffusion processes.
\newblock {\em Communications in Nonlinear Science and Numerical Simulation}, 106:106095, 2022.

\bibitem{tong2017modelling}
Z.~Tong.
\newblock Modelling VIX and VIX derivatives with reducible diffusions.
\newblock {\em International Journal of Bonds and Derivatives}, 3(2):153--175, 2017.

\bibitem{wood2004chain}
J.~A. Wood.
\newblock The chain rule for matrix exponential functions.
\newblock {\em The College Mathematics Journal}, 35(3):220--222, 2004.

\bibitem{yao2006regime}
D.~D. Yao, Q.~Zhang, and X.~Y. Zhou.
\newblock Stochastic processes, optimization, and control theory: Applications in financial engineering,
queueing networks, and manufacturing systems.
\newblock In {\em Stochastic Processes, Optimization, and Control Theory}, pages 281--300. Springer, Boston, 2006.

\bibitem{yuan2004convergence}
C.~Yuan and X.~Mao.
\newblock Convergence of the Euler--Maruyama method for stochastic differential equations with Markovian switching.
\newblock {\em Mathematics and Computers in Simulation}, 64(2):223--235, 2004.

\bibitem{zhou2003ito}
H.~Zhou.
\newblock It\^{o} conditional moment generator and the estimation of short-rate processes.
\newblock {\em Journal of Financial Econometrics}, 1(2):250--271, 2003.

\bibitem{zhu2012new}
S.-P. Zhu, A.~Badran, and X.~Lu.
\newblock A new exact solution for pricing European options in a two-state regime-switching economy.
\newblock {\em Computers \& Mathematics with Applications}, 64:2744--2755, 2012.

\end{thebibliography}
\end{document}